\documentclass[10pt,twocolumn,letterpaper]{article}

\usepackage{cvpr}              
\usepackage{graphicx}
\usepackage{amsmath}
\usepackage{amssymb}
\usepackage{booktabs}
\usepackage{makecell}  
\usepackage{multirow}
\usepackage{textcomp}
\usepackage{comment}
\usepackage[dvipsnames]{xcolor}

\definecolor{cvprblue}{rgb}{0.21,0.49,0.74}
\usepackage[pagebackref,breaklinks,colorlinks,citecolor=cvprblue]{hyperref}

\def\paperID{7631} 
\def\confName{CVPR}
\def\confYear{2024}

\title{EyeBench: A Call for More Rigorous Evaluation of Retinal Image Enhancement}

\author{Wenhui Zhu$^{1*}$\enspace Xuanzhao Dong $^{1*}$ \enspace  Xin Li$^{1*}$\enspace Yujian Xiong$^{1*}$ \enspace Xiwen Chen $^{2}$ \\
\enspace Peijie Qiu$^{3}$ \enspace Vamsi Krishna Vasa$^{1}$ \enspace Zhangsihao Yang$^{1}$ \enspace Yi Su$^{4}$ \enspace Oana Dumitrascu$^{5}$ \enspace Yalin Wang$^{1}$ \\
$^{1}$ Arizona State University,  
$^{2}$ Clemson University, 
$^{3}$ Washington University in St. Louis, \\
$^{4}$ Banner Alzheimer's Institute, 
$^{5}$ Mayo Clinic
}

\begin{document}

\twocolumn[{%
\renewcommand\twocolumn[1][]{#1}%
\maketitle
\begin{center}
    \centering
    \captionsetup{type=figure}
    \includegraphics[width=1.0\textwidth]{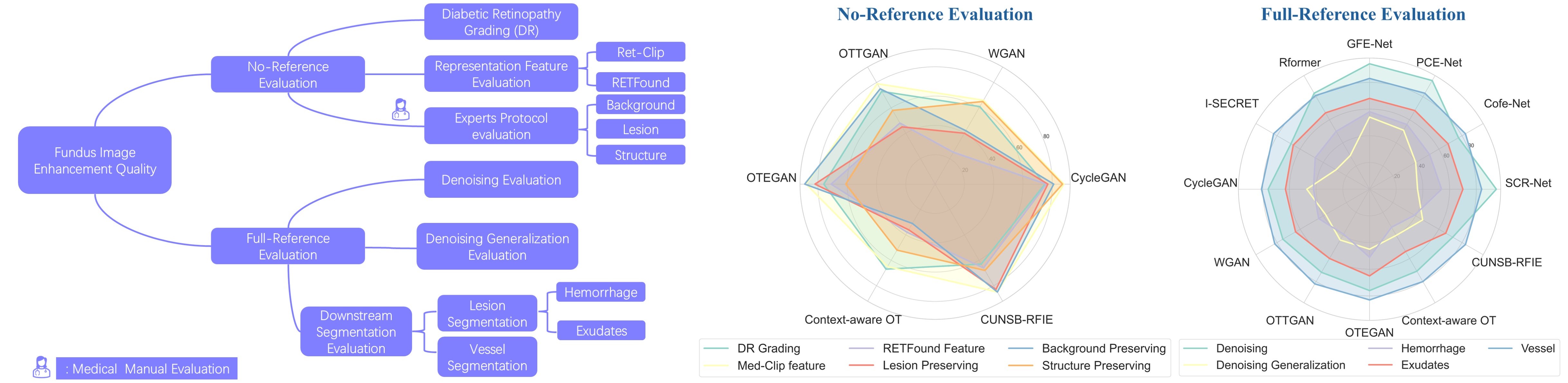}
    \captionof{figure}{ 
    \textbf{Overview of EyeBench.} We introduce EyeBench, a systematic and rigorous benchmark for evaluating retinal image enhancement models. Our evaluation pipeline comprehensively assesses fundus image enhancement quality through both No-Reference and Full-Reference aspects, facilitating a multi-dimensional evaluation. For each aspect, we design a \textbf{distribution-aligned} dataset to ensure fair and clinically meaningful comparisons. 
    Additionally, we include \textbf{clinically consistent} downstream tasks to quantify models' ability in denoising generalization and downstream preserving. Our benchmark also incorporates \textbf{medical experts guided annotations}, adhering to expert protocols, and we statistically validate that EyeBench results \textbf{aligned well with clinic preference assessment}. Finally, we highlight current challenges to inform future development. EyeBench can provide multiple insights from multiple perspectives. 
    }
    \label{banner}
\end{center}%
}]

\def\thefootnote{*}\footnotetext{These authors contributed equally to this paper.}


\begin{abstract}
Over the past decade, generative models have achieved significant success in enhancement fundus images.
However, the evaluation of these models still presents a considerable challenge. A comprehensive evaluation benchmark for fundus image enhancement is indispensable for three main reasons: 1) The existing denoising metrics (e.g., PSNR, SSIM) are hardly to extend to downstream real-world clinical research (e.g., lesion preserving, Vessel morphology consistency). 2) There is a lack of comprehensive evaluation for both paired and unpaired enhancement methods, along with the need for  expert protocols to accurately assess clinical value.  3) An ideal evaluation system should provide insights to inform future developments of fundus image enhancement. To this end, we propose a novel comprehensive benchmark, \textbf{EyeBench}, to provide insights that align enhancement models with clinical needs, offering a foundation for future work to improve the clinical relevance and applicability of generative models for fundus image enhancement. EyeBench has three appealing properties: \textbf{1)  multi-dimensional clinical alignment downstream evaluation:} In addition to evaluating the enhancement task, we provide several clinically significant downstream tasks for fundus images, including vessel segmentation, DR grading, denoising generalization, and lesion segmentation.  \textbf{2) Medical expert-guided evaluation design:} We introduce a novel dataset that facilitates comprehensive and fair comparisons between paired and unpaired methods and includes a manual evaluation protocol by medical experts (e.g., the ratio of lesion structure changed, background-color changed, and extra structures generated). \textbf{3) Valuable insights:} Our benchmark study provides a comprehensive and rigorous evaluation of existing methods across different downstream tasks, assisting medical experts in making informed choices. Additionally, we offer analysis of the challenges faced by existing methods, which would shine a light for the further design of generative models for fundus image enhancement.The code is available at \url{https://github.com/Retinal-Research/EyeBench}

\end{abstract}
\section{Introduction}
Non-mydriatic retinal color fundus photography (CFP) is widely used in various fundus disease analyses due to the advantage of not requiring pupillary dilation~\cite{deeplearning1,deeplearning2,deeplearning3,deeplearning4,deeplearning5,deeplearning6}. However, it commonly suffers low quality due to artifacts, uneven illumination, deficient ocular media transparency, poor focus, or inappropriate imaging~\cite{shen2020modeling,fu2019evaluation}. Recently, fundus image enhancement has witnessed significant advancements with the rapid development of the generative model. 
Since these models are not constrained by the need for paired data, a growing number of unpaired image enhancement models have been developed, showing performance comparable to paired methods~\cite{wang2022optimal,zhu2023optimal, zhu2023otre,dong2024cunsb,vasa2024context}. At the same time, due to the difficulty of collecting paired noisy and high-quality images, medical experts now show a stronger preference for these unpaired methods. However, the existing evaluation metrics for fundus image enhancement still comply with the supervised denoising task where the low-high quality fundus image pair synthesis by adding known noises (e.g., Gaussian blur, white noise) into real-world high-quality images. This evaluation heavily relies on conventional metrics such as SSIM and PSNR, which fall short of thoroughly assessing the denoising capabilities and similarity between the latent representations of enhanced images and real high-quality images. Moreover, enhancement evaluation alone does not meet clinical requirements, and a rigorous evaluation framework is needed for both unpaired and paired methods to ensure comprehensive assessment. In this paper, we introduce EyeBench, a comprehensive and rigorous benchmark for evaluating fundus image enhancement methods, which includes the multi-dimensional clinical alignment downstream
evaluation, medical expert-guided evaluation design, and valuable insights. 

First, our benchmark introduces a set of downstream tasks to assess enhanced fundus images, breaking down enhancement quality into clinical preferences, specifically focusing on preserving vessels, disease grading, and lesion structures. We train existing enhancement methods within a standardized framework and apply them to improve fundus image quality for each downstream task. These enhanced images are then processed through respective evaluation workflows for further analysis. As shown in Fig.~\ref{banner}, the downstream tasks include enhancement generalization, vessel segmentation, lesion segmentation, representation, and diabetic retinopathy grading (DR grading). These tasks assess the discrepancies between the generated masks, labels, or representations of the enhanced images and the ground truth or high-quality images. This allows us to determine if vessel structures remain intact and lesion areas are preserved. In addition, the proposed evaluation assesses the enhancement performance and improves the credibility of different enhancement methods for clinical applications.

Second, we annotate unusable images and resample the labels for disease severity levels for each sub-set following guidance from medical experts. To facilitate a more rigorous comparison between paired and unpaired methods under real-world (no-reference) and synthetic (full-reference) noise, we propose a new dataset specifically designed to include dedicated training and testing sets for both paired and unpaired methods under full-reference conditions, allowing for the evaluation of denoising and various downstream tasks. Under no-reference conditions, we restructure the training and testing sets to assess the performance of unpaired methods. Furthermore, we introduce an expert manual evaluation protocol, as shown in Fig.~\ref{banner}, to align with clinical preference by assessing enhanced images. We also conducted a statistical analysis of expert annotation evaluation and EyeBench to validate the necessity of multi-dimensional evaluation.

Third, our multi-dimensional evaluation result (see Fig.~\ref{banner}) will assist medical experts in selecting the appropriate enhancement methods to improve the reliability of subsequent diagnoses and analyses. Specifically for clinically valuable unpaired methods, we provide a detailed analysis of denoising generalization. Additionally, we offer comprehensive analyses and insights into the challenges that these existing methods face and insights for future works.


\section{Existing Methods} \label{Sec: methods}
We aim to explore current image-denoising methods, focusing on paired and unpaired training approaches. To facilitate this discussion, we let $\mathbf{X}_i$, and $\mathbf{Y}_i$ represent low-quality and high-quality images, respectively, with corresponding distribution $\mathbb{P}_{\mathbf{X}_i}$ and $\mathbb{P}_{\mathbf{Y}_i}$, where the disjoint set index $i \in \{1,2 \}$. For all paired methods outlined in Sec.~\ref{Sec:supervised}, we focus on data pairs $(\mathbf{x}_1,\mathbf{y}_1)$ such that $\mathbf{x}_1\sim \mathbb{P}_{\mathbf{X}_1}$ and $\mathbf{y}_1 \sim \mathbb{P}_{\mathbf{Y}_1}$. In contrast, for unpaired methods discussed in Sec.~\ref{Sec:unsupervised}, the data is represented as $\mathbf{x}_1\sim \mathbb{P}_{\mathbf{X}_1}, \mathbf{y}_2 \sim \mathbb{P}_{\mathbf{Y}_2}$, ensuring that no paired information is available.

\subsection{Paired Methods}\label{Sec:supervised}
Leveraging pairs of degraded and clean images, denoted as $(\mathbf{x}_1,\mathbf{y}_1)$, paired methods in retinal fundus image enhancement can be uniformly expressed as:
\begin{equation}\label{eq:paired-methods}
    \hat{\mathbf{y}}_1 = f_\theta(\mathbf{x}_1)
\end{equation}
\noindent Here, $f_\theta$ represents the denoising network, which utilizes a degradation model to simulate noise in fundus images and applies various neural network architectures to restore image quality. In methods such as \textit{SCR-Net}~\cite{li2022structure}, \textit{Cofe-Net}~\cite{shen2020modeling}, \textit{PCE-Net}~\cite{10.1007/978-3-031-16434-7_49} and \textit{GFE-Net}~\cite{li2023generic}, $\hat{\mathbf{y}}_1$ is modeled using a Variational autoencoder (VAE), incorporating additional information (e.g., high-frequency details, latent retinal structure, artifacts, and the Laplacian Pyramid Features) to regularize the denoising process. In contrast, \textit{RFormer}~\cite{deng2022rformer}, $f_\theta$ employs a transformer-based generator, where $f_\theta$ focuses on capturing long-range dependencies present in $\mathbf{x}_1$. 
Notably, \textit{I-SECRET}~\cite{i-secret} leverages a semi-supervised approach to optimize $f_\theta$. In the initial two training phases, paired images are utilized to ensure that $f_\theta$ can preserve structural details and maintain pixel-wise alignment. Subsequently, adversarial learning is applied in an unpaired training setting, where $f_\theta$ functions as an optimized generator. For the sake of consistency in our experiments, we categorize this model as a paired method.

\subsection{Unpaied Methods} \label{Sec:unsupervised}
Unpaired methods in retinal image denoising can be broadly categorized into two main approaches:  \textbf{GAN-based} and \textbf{SDE-based} methods, employing the generative models (e.g., GAN~\cite{goodfellow2020generative}, Diffusion~\cite{ho2020denoising,song2020denoising}, Gradient flow~\cite{song2019generative}). Since collecting paired clean and noisy images from the real world is challenging, the prevailing approach frames the denoising task as a style transfer problem.

\noindent\textbf{GAN-based model}. The adversarial learning strategy enhances GAN-based models in generating realistic retinal images with detailed structures. The typical adversarial objective is formulated as follows:
\begin{equation}\label{eq:gan-typical}
\begin{split}
    \min_{G_{\mathbf{X}_1}} \max_{D_{\mathbf{Y}_2}} \mathcal{L}  &:= \mathbb{E}_{\mathbf{y}_2}[\log D_{\mathbf{Y}_2}(\mathbf{y}_2)] \\
   & + \mathbb{E}_{\mathbf{x}_1 }[\log(1 - D_{\mathbf{Y}_2}(G_{\mathbf{X}_1}(\mathbf{x}_1)))]
\end{split}
\end{equation}
\noindent Here, the generator $G_{\mathbf{X}_1}$ and discriminator $D_{\mathbf{Y}_2}$ work in opposition, seeking to converge toward a Nash Equilibrium. 

\noindent\textit{CycleGAN}~\cite{cyclegan} addresses the limitation of requiring paired data by duplicating the GAN structure. It incorporates cycle consistency and identity regularization to support two-way transformations. 
Specifically, the cycle consistency loss enforces bidirectional mapping, improving alignment and coherence in the generated images. However, these additional structures increase computational complexity and may result in suboptimal performance (e.g., mode collapse, artifacts), especially when dealing with images that exhibit multimodal distributions~\cite{salmona2022can}.

\noindent\textit{Wasserstein-GAN} (WGAN)~\cite{arjovsky2017wasserstein,gulrajani2017improved} is a widely recognized model rooted in OT theory. Instead of solving the primal OT problem directly, WGAN leverages the Kantorovich-Rubinstein duality~\cite{villani2009optimal} to approximate the Wasserstein distance in a computationally feasible manner. The objective function is given by:
\begin{equation}\label{eq:wgan-basic}
    \begin{split}
        \min_{G_{\mathbf{X}_1}} \max_{D_{\mathbf{Y}_2}} \mathcal{L} &:= \mathbb{E}_{\mathbf{y}_2 }[D_{\mathbf{Y}_2}(\mathbf{y}_2)] 
        - \mathbb{E}_{\mathbf{x}_1 } [D_{\mathbf{Y}_2}(G_{\mathbf{X}_1}(\mathbf{x}_1))]
    \end{split}
\end{equation}
\noindent Here, the generator $G_{\mathbf{X}_1}$ learns to map $\mathbb{P}_{\mathbf{X}_1}$ to $\mathbb{P}_{\mathbf{Y}_2}$ by minimizing the Wasserstein distance between them. The discriminator $D_{\mathbf{Y}_2}$ maximizes the difference in continuous scores assigned to the real high-quality images and the synthetic images generated by $G_{\mathbf{X}_1}$. By providing feedback on the quality of the generated images, the discriminator guides the generator toward the optimal mapping.

In contrast to WGAN, which approximates the Wasserstein distance indirectly, \textit{OTT-GAN}~\cite{wang2022optimal} directly solves the Monge's Optimal Transport problem using an adversarial training strategy. The objective function is expressed as:
\begin{equation} \label{eq:ott-basic}
    \begin{split}
        \max_{G_{\mathbf{X}_1}} \min_{D_{\mathbf{Y}_2}} \mathcal{L}:= \mathbb{E}_{\mathbf{x}_1} [ C( \mathbf{x}_1, G_\theta (\mathbf{x}_1) )] 
        +\lambda \mathbf{W}_1(\mathbb{P}_{\mathbf{Y}_2}, \mathbb{P}_{G_{\mathbf{X}_1}(\mathbf{x}_1)} )
    \end{split}
\end{equation}
\noindent Here, the cost function $C$ is defined as mean square error (MSE), and the method in Eq.~\ref{eq:wgan-basic} is employed to approximate the Wasserstein distance, denoted as $\mathbf{W}_1$. Building on \textit{OTT-GAN}, \textit{OTE} and \textit{OTRE}~\cite{zhu2023optimal,zhu2023otre} incorporates the Multi-Scale Structural Similarity Index Measure (MS-SSIM)~\cite{wang2003multiscale,brunet2011mathematical} as the cost function, along with identity regularization, which improve the preservation of structural details during image translation. To further enhance contextual preservation during denoising, \textit{Context-aware OT}~\cite{vasa2024context} extends beyond pixel-based costs. It leverages a pretrained VGG~\cite{mechrez2018contextual} network to capture the earth mover's distance in the feature space, thereby improving perceptual similarity between the input and generated images.

\noindent\textbf{SDE-based model}. \textit{CUNSB-RFIE}~\cite{dong2024cunsb} seeks to identify the Schr\"{o}dinger Bridge (SB), denoted as $\mathbb{Q}^{SB}$, between $\mathbb{P}_{\mathbf{X}_1}$ and $\mathbb{P}_{\mathbf{Y}_2}$. This approach enables a smooth and probabilistically consistent transformation between image distributions, but it results in a loss of high-frequency information during the iterative training process. The main objective function for an arbitrary step $t_i$ is expressed as:
\begin{equation}\label{eq:CUNSB-RFIE}
\begin{split}
    &\min_{\phi} \mathbb{L}(\phi,t_i) := \mathbb{L}_{Adv}(\phi,t_i) + \lambda_{SB} \mathbb{L}_{SB}(\phi,t_i) 
\end{split}
\end{equation}
\noindent Here, $\phi$ parameterizes the generator $G_{\mathbf{X}_1}$ at step $t_i$. The term $\mathbb{L}_{Adv}$ modifies the KL-divergence between the synthetic high-quality image distribution and the ground-truth distribution $\mathbb{P}_{\mathbf{Y}_2}$, while $ \mathbb{L}_{SB}$ approximates the solution to the entropy-regularized optimal transport problem. Consequently, the final static solution shares the same marginal distributions as $\mathbb{Q}^{SB}$~\cite{tong2023improving}. 

\begin{figure*}[h]
  \centering
  \includegraphics[width=0.9\textwidth]{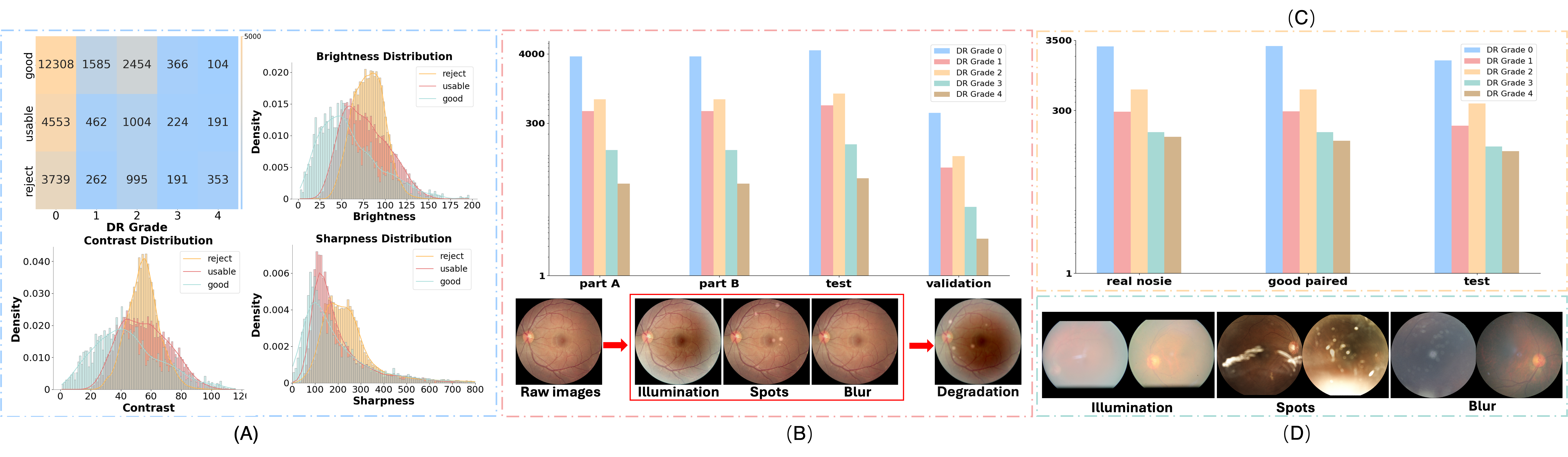}  
  \caption{ (A) highlights attribute distributions (i.e., brightness, contrast, sharpness) and diabetic retinopathy (DR) grades across quality categories (i.e., good, usable, and reject). (B) illustrates histograms for the training (i.e., part A and part B), testing, and validation datasets used in \textbf{Full-Reference} evaluations after resampling, with the workflow of degradation algorithms outlined below. (C) shows histograms for real-world \textbf{No-Reference} experiments after resampling. (D) presents samples to be overprocessed.
  }
  \label{fig:overall_ana}
\end{figure*}

\section{Clinic Experts Guided Data Annotation}

Our dataset was sourced from the EyePACS initiative~\cite{diabetic-retinopathy-detection}, with quality annotations derived from the EyeQ dataset~\cite{fu2019evaluation}. We collected 28,791 color fundus images, which were categorized into three quality levels: good, usable, and reject. Additionally, each image was annotated with diabetic retinopathy (DR) severity labels across five levels (0, 1, 2, 3, and 4), where higher values indicate greater severity of DR.
Analyses of attribute distributions (i.e., brightness, contrast, and sharpness) and DR grade across quality categories are presented in Fig.~\ref{fig:overall_ana}(A). The results reveal notable attribute discrepancies across quality categories, with the distributions for good and usable quality images closely aligned. In contrast, the reject category shows significant differences (e.g., a higher prevalence of high-acutance images). Additionally, DR labels are imbalanced, with labels 0 and 2 being more frequent and severe DR cases (labels 3 and 4) relatively underrepresented. Finally, as illustrated in Fig.~\ref{fig:overall_ana}(D), some rejected and usable quality examples tend to be overprocessed, compromising their diagnostic quality and clinical practicability. We re-selected good and usable quality images and applied ratio-preserving resampling to maintain lesion information alignment based on medical expert guidance. Given the low prevalence of severe DR cases, achieving a balanced sample across DR grades is challenging. Therefore, we retained the natural distribution of DR labels across subsets to better represent real-world clinical scenarios and align with objective engineering principles.

\noindent \textbf{Full-Reference Evaluation Dataset.} A total of 16,817 good-quality images were used here, with 10,000 images for training, 600 for validation, and 6,217 for testing, as detailed in Fig.~\ref{fig:overall_ana}(B). To support the experiments, all good-quality images were degraded using the algorithms outlined in ~\cite{shen2020modeling}, which simulate the combinations of illumination, spot artifacts, and blurring. Additionally, the training set was split into two disjoint subsets of 5,000 images each, referred to as $A$ and $B$, with synthesized noisy paired subsets $A^{\ast}$ and $B^{\ast}$, which were used to train the paired ($A^{\ast}$ to $A$) and unpaired ($A^{\ast}$ to $B$) methods.

\noindent \textbf{No-Reference Evaluation Dataset.}
As shown in Fig.~\ref{fig:overall_ana}(C), A total of 6,434 usable-quality images were included in this study, and all usable-quality images were resampled, resulting in 4,000 real-world noisy images (i.e., real noise) for training and 2,434 testing images. Additionally, 4,000 unpaired good-quality images were resampled from the original set of 10,000 good-quality training images based on the DR label, ensuring the experiment followed the unpaired training scheme.

\begin{table*}[!t]
\centering
\caption{Performance comparison of denoising evaluation in Full-Reference quality assessment experiments. The best performance in each column is highlighted in bold, with the second-best underlined. Visualization results refer to the \textcolor{red}{Appendix C}.}
\tiny
\resizebox{0.8\textwidth}{!}{%
\begin{tabular}{cccccccc}
\toprule
\multirow{2}{*}{} & \multirow{2}{*}{\textbf{Method}} & \multicolumn{2}{c}{\textbf{EyeQ}} & \multicolumn{2}{c}{\textbf{IDRID}} & \multicolumn{2}{c}{\textbf{DRIVE}} \\ \cmidrule(l){3-8} 
                                          &                                  & \textbf{SSIM} $\uparrow$   & \textbf{PSNR} $\uparrow$   & \textbf{SSIM} $\uparrow$   & \textbf{PSNR} $\uparrow$   & \textbf{SSIM} $\uparrow$   & \textbf{PSNR} $\uparrow$   \\ \midrule
\multirow{5}{*}{\textit{Paired Methods}} & SCR-Net~\cite{li2022structure}   & \textbf{0.9606} & 29.698 & 0.6425 & 18.920 & \textbf{0.6824} & 23.280 \\      
                                         & Cofe-Net~\cite{shen2020modeling} & 0.9408           & 24.907           & 0.7397            & 20.058            & 0.6671            & 21.774            \\
                                         & PCE-Net~\cite{10.1007/978-3-031-16434-7_49} & 0.9487           & \textbf{29.895}           & \underline{0.7764}            & \underline{23.201}           & 0.6704            & 24.041           \\
                                         & GFE-Net~\cite{li2023generic}     & \underline{0.9554}           & \underline{29.719}           & \textbf{0.7935}            & \textbf{25.012}           & \underline{0.6793}            & \underline{23.786} \\
                                         &RFormer~\cite{deng2022rformer}
                                          & 0.9260   & 27.163   & 0.5963   & 18.433   & 0.6311   & 22.172\\
                                          & I-SECRET~\cite{i-secret}        & 0.9051 & 23.483 & 0.7157 & 20.173 & 0.5727 & 18.803 \\
                                          \midrule
\multirow{7}{*}{\textit{Unpaired Methods}} 
                                         & CycleGAN~\cite{cyclegan}         & \underline{0.9313}           &\textbf{25.076}           & \textbf{0.7668}            & \textbf{22.511}           & \textbf{0.6681}            & \textbf{22.686}  \\
                                         & WGAN~\cite{gulrajani2017improved} & 0.9266          & 24.793           & 0.7316           & 21.325           & 0.6431            & 20.408 \\
                                         & OTTGAN~\cite{wang2022optimal}    & 0.9275           & 24.065           & 0.7509           & 22.131           & 0.6635            & 21.938 \\
                                         & OTEGAN~\cite{zhu2023optimal}     & \textbf{0.9392}           & \underline{24.812}           & 0.7624           & 22.272           & 0.6642            & 22.183 \\
                                         & Context-aware OT~\cite{vasa2024context} & 0.9144           & 24.088           & 0.7338            & 21.790            & 0.6407            & 21.389 \\
                                         & CUNSB-RFIE~\cite{dong2024cunsb}  & 0.9121           & 24.242           & \underline{0.7651}           & \underline{22.448}           & \underline{0.6659}            & \underline{22.510} \\
\bottomrule
\end{tabular}%
}
\label{tb:deg-exp}
\end{table*}

\begin{table*}[!t]
    \centering
    \caption{
    Performance comparison of vessel and lesion (EX and HE) segmentation in Full-Reference quality assessment experiments. The best performance in each column is highlighted in bold, with the second-best underlined. For visualization results, refer to the \textcolor{red}{Appendix C}.}
    \tiny
    \resizebox{0.9\textwidth}{!}{%
    \begin{tabular}{lcccc|ccc|ccc}
        \toprule
         \multirow{2}[3]{*}{Method} & \multicolumn{4}{c}{Vessel Segmentation} & \multicolumn{3}{c}{EX} & \multicolumn{3}{c}{HE} \\ 
         \cmidrule(lr){2-5}  \cmidrule(lr){6-8}  \cmidrule(lr){9-11}
          
         & AUC $\uparrow$ & PR $\uparrow$ & F1 Score $\uparrow$ & SP $\uparrow$ & AUC  & PR & F1 Score  & AUC & PR & F1 Score \\ \midrule

        SCR-Net~\cite{li2022structure}   & \textbf{0.9227} & \textbf{0.7783} & \textbf{0.7000} & 0.9787 & \textbf{0.9683} & \textbf{0.6041} & \textbf{0.5556} & 0.9377 & 0.3213 & 0.3725\\ 
        cofe-Net~\cite{shen2020modeling} & \underline{0.9188} & \underline{0.7698} & 0.6895 & 0.9801 & 0.9623 & 0.5620 & 0.5349 & 0.9302 & 0.3152 & 0.3281\\
        PCE-Net~\cite{10.1007/978-3-031-16434-7_49} & 0.9146 & 0.7616 & 0.6790 & \textbf{0.9814} & \underline{0.9667} & \underline{0.5876} & 0.5066 & \underline{0.9545} & \underline{0.3639} & \underline{0.3736}\\
        GFE-Net~\cite{li2023generic} & 0.9175 & 0.7669 & 0.6832 & \textbf{0.9814} & 0.9560 & 0.5548 & \underline{0.5380} & \textbf{0.9577} & \textbf{0.4113} & \textbf{0.3751}\\
        RFormer~\cite{deng2022rformer} & 0.8990 & 0.7239 & 0.6374 & \underline{0.9806} & 0.9626 & 0.5593 & 0.4692 & 0.9207 & 0.2677 & 0.3136 \\
        I-SECRET~\cite{i-secret} & 0.9181 & 0.7662 & 0.6838 & 0.9802 & 0.9613 & 0.5424 & 0.4825 & 0.9028 & 0.2629 & 0.2642 \\
    \midrule
        CycleGAN~\cite{cyclegan} & 0.9015 & 0.7278 & 0.6462 & \underline{0.9801}  & 0.9447& 0.4843 & 0.4790 & 0.8970 & 0.1624 & 0.2227\\
        WGAN~\cite{gulrajani2017improved} & 0.9081 & 0.7494 & 0.6768 & 0.9764  & 0.9522 & 0.4942 & 0.4859 & \underline{0.8990} & \underline{0.1847} & \underline{0.2476}\\
        OTTGAN~\cite{wang2022optimal} & 0.9034 & 0.7400 & 0.6609 & \textbf{0.9812}  & 0.9492 & 0.4214 & 0.4365 & 0.8179 & 0.1448 & 0.2233 \\
        OTEGAN~\cite{zhu2023optimal} & \underline{0.9156} & \textbf{0.7678} & \textbf{0.6919} & 0.9797 & \underline{0.9562} & \underline{0.5191} & \underline{0.4868} & \textbf{0.9359} & \textbf{0.2800}& \textbf{0.3165} \\
        Context-aware OT~\cite{vasa2024context} & 0.8871 & 0.7077 & 0.6377 & 0.9791 & 0.9305 & 0.3318 & 0.3707 & 0.8091 & 0.0646 & 0.1184 \\
        CUNSB-RFIE~\cite{dong2024cunsb}  & \textbf{0.9163} & \underline{0.7626} & \underline{0.6872} & 0.9784 & \textbf{0.9572}& \textbf{0.5381} & \textbf{0.4883} & 0.8488 & 0.1489 & 0.1893 \\
        \bottomrule
    \end{tabular}}
    \label{tab-seg}
    \vspace{-0.3cm}
\end{table*}

\begin{table*}[!t]
    \centering
    \caption{Performance comparison with unpaired baselines in No-Reference quality assessment task. The best performance in each column is highlighted in bold, and the second-best is underlined. Visualization results refer to \textcolor{red}{Appendix C}.}
    \tiny
    \resizebox{0.9\textwidth}{!}{%
    \begin{tabular}{lcccc|cc|ccc}
        \toprule
         \multirow{2}[3]{*}{Method} & \multicolumn{4}{c}{DR grading} & \multicolumn{2}{c}{Representation Feature} & \multicolumn{3}{c}{Experts Protocol Evaluation} \\ 
         \cmidrule(lr){2-5}  \cmidrule(lr){6-7}  \cmidrule(lr){8-10}
          
         & ACC $\uparrow$ & Kappa $\uparrow$ & F1 Score $\uparrow$ & AUC $\uparrow$ & FID-Retfound~\cite{zhou2023foundation}$\downarrow$  & FID-Clip~\cite{du2024ret} $\downarrow$ & LPR $\uparrow$ & BPR $\uparrow$ & SPR $\uparrow$\\ \midrule
         
        CycleGAN~\cite{cyclegan}                & \textbf{0.7588} & \underline{0.6006} & \underline{0.7180} & \underline{0.9251}  & \textbf{23.778}& \underline{11.530}  & \underline{0.7707} & 0.8153 & \textbf{0.8726}\\
        
        WGAN~\cite{gulrajani2017improved}       & 0.6446 & 0.3123 & 0.6156 & 0.8874  & 74.885 & 33.076  & 0.4076 & 0.4204 & 0.6561\\
        
        OTTGAN~\cite{wang2022optimal}           & 0.7440 & 0.5688 & 0.7037 & 0.9247  & 51.201 & 20.505 & 0.4586 & 0.7580 & 0.5860 \\
        
        OTEGAN~\cite{zhu2023optimal}            & \underline{0.7539} & \textbf{0.6433} & \textbf{0.7228} & \textbf{0.9326} & \underline{28.987} & \textbf{11.114} & \textbf{0.8280} & \textbf{0.8981} & 0.6178 \\
        
        Context-aware OT~\cite{vasa2024context} & 0.7301 & 0.3811 & 0.6662 & 0.9112 & 61.429 & 34.456 & 0.3566 & 0.3121 & 0.5159 \\
        
        CUNSB-RFIE~\cite{dong2024cunsb}         & 0.6565 & 0.3674 & 0.6341 & 0.8927 & 33.047 & 14.827  & \textbf{0.8280} & \underline{0.8535} & \underline{0.6879} \\
        
        \bottomrule
    \end{tabular}}
    \label{tab-noref}
    \vspace{-0.3cm}
\end{table*}

\section{Experiments}


\subsection{Full-Reference Quality Assessment Experiments} \label{section-4.1}
For full-reference assessment, we used the previously synthesized Full-Reference Assessment Dataset. We strictly followed the training configurations for paired and unpaired methods. For the unpaired method, synthetic low-quality images from training set $A$ (i.e., $A^{\ast}$) were used as input images, while high-quality images from training set $B$ served as the clean reference images. For the paired method, we performed supervised training using low-high-quality image pairs from the training set $A$ (i.e., $A^{\ast}$ and $A$). All models were trained with the parameter report in the original paper. For two segmentation tasks, we trained a vanilla U-Net~\cite{ronneberger2015unet} model from scratch.
The following baselines were considered in this evaluation: \textit{Paired algorithms}: SCR-Net~\cite{li2022structure}, Cofe-Net~\cite{shen2020modeling}, PCE-Net~\cite{10.1007/978-3-031-16434-7_49}, GFE-Net~\cite{li2023generic}, RFormer~\cite{deng2022rformer}, \textit{Unpaired algorithms}: I-SECRET~\cite{i-secret}, CycleGAN~\cite{cyclegan}, WGAN~\cite{gulrajani2017improved}, OTTGAN~\cite{wang2022optimal}, OTEGAN~\cite{zhu2023optimal}, Context-aware OT~\cite{vasa2024context}, CUNSB-RFIE~\cite{dong2024cunsb}.
We generated enhanced images for the trained models separately for the downstream evaluations. Refer to the \textcolor{red}{Appendix A} for more details.

\noindent \textbf{Denoising Evaluation.} We input the noisy images from the Full-Reference testing set into the trained models to generate enhanced images. These enhanced low-quality images were then evaluated using the Peak Signal-to-Noise Ratio (PSNR) and Structural Similarity Index Measure (SSIM).

\noindent \textbf{Denoising generalization Evaluation.} To evaluate denoising generalization, we degraded high-quality images following the same degradation algorithm to synthesize the low-quality images for DRIVE~\cite{drive} and IDRID~\cite{idrid}. Similarly, we fed these enhanced images into the trained model to calculate the PSNR and SSIM between the enhanced and original images (treated as high-quality images).

\noindent \textbf{Vessel Segmentation}. The vessel segmentation task is performed using the DRIVE dataset, which includes annotated masks to further evaluate the ability to preserve blood vessel structures during denoising. We follow the official split, resulting in 20 subjects each in the training and testing sets. We use the enhanced images as training and testing images generated from the generalization evaluation. The vessel segmentation task is evaluated using the Area Under the ROC Curve (AUC), the Area under the Precision-Recall Curve (PR), F1 Score, and Specificity (SP).

\noindent \textbf{Lesion Segmentation}. We use the segmentation masks provided with the IDRID dataset. Since the downstream segmentation tasks were trained and tested solely on enhanced images(obtained from the generalization task), without additional preprocessing or enhancements, we focused only on larger, easier-to-train lesion types, including Hard Exudates (EX) and Hemorrhages (HE). The training set includes 54 subjects, while the testing set includes 27 subjects. Performance is measured using AUC, PR, and F1 score.

\subsection{No-Reference Quality Assessment Experiments}
Evaluating enhancement quality without ground-truth clean images presents a particular challenge for paired methods. Therefore, we focused on unpaired methods to assess real-world denoising capabilities. For this evaluation, we trained the unpaired method using the No-Reference Assessment Dataset, processing 2,434 low-quality testing images to generate enhanced images. These enhanced images were then used in various downstream evaluations, including DR grading, representation feature analysis, and expert assessment by medical professionals. A no-reference quality assessment was conducted on the following baselines: CycleGAN~\cite{cyclegan}, WGAN~\cite{gulrajani2017improved}, OTTGAN~\cite{wang2022optimal}, OTEGAN~\cite{zhu2023optimal}, Context-aware OT~\cite{vasa2024context}, and CUNSB-RFIE~\cite{dong2024cunsb}.
All models were trained using the parameters followed in the original papers. Refer to the \textcolor{red}{Appendix B} for more details.

\noindent\textbf{DR grading.} We trained an NN-MobileNet model~\cite{deeplearning1} for the DR grading task using real-world high-quality images. The enhanced test images are used with the trained NN-MobileNet to infer DR grading classification. Enhancement performance is evaluated based on classification accuracy (ACC), kappa score, F1 score, and AUC. This evaluation primarily aims to assess whether the denoising model disrupts lesion distribution, potentially leading to inconsistencies with the original DR grading labels.

\noindent\textbf{Representation Feature Evaluation.} We employed two fundus image-based foundation models (Retfound~\cite{zhou2023foundation} and Ret-clip~\cite{du2024ret}) to calculate the Fréchet inception distance (FID) between enhanced and real-world high-quality image feature representation, referred to as \textit{FID-Retfound} and \textit{FID-Clip}. \textit{FID-Retfound} measures the preservation of disease-related information, while \textit{FID-Clip} assesses the similarity of spatial structures and continuous features. 

\noindent\textbf{Experts Annotation Evaluation.}
\begin{figure}
    \centering
    \includegraphics[width=1.0\linewidth]{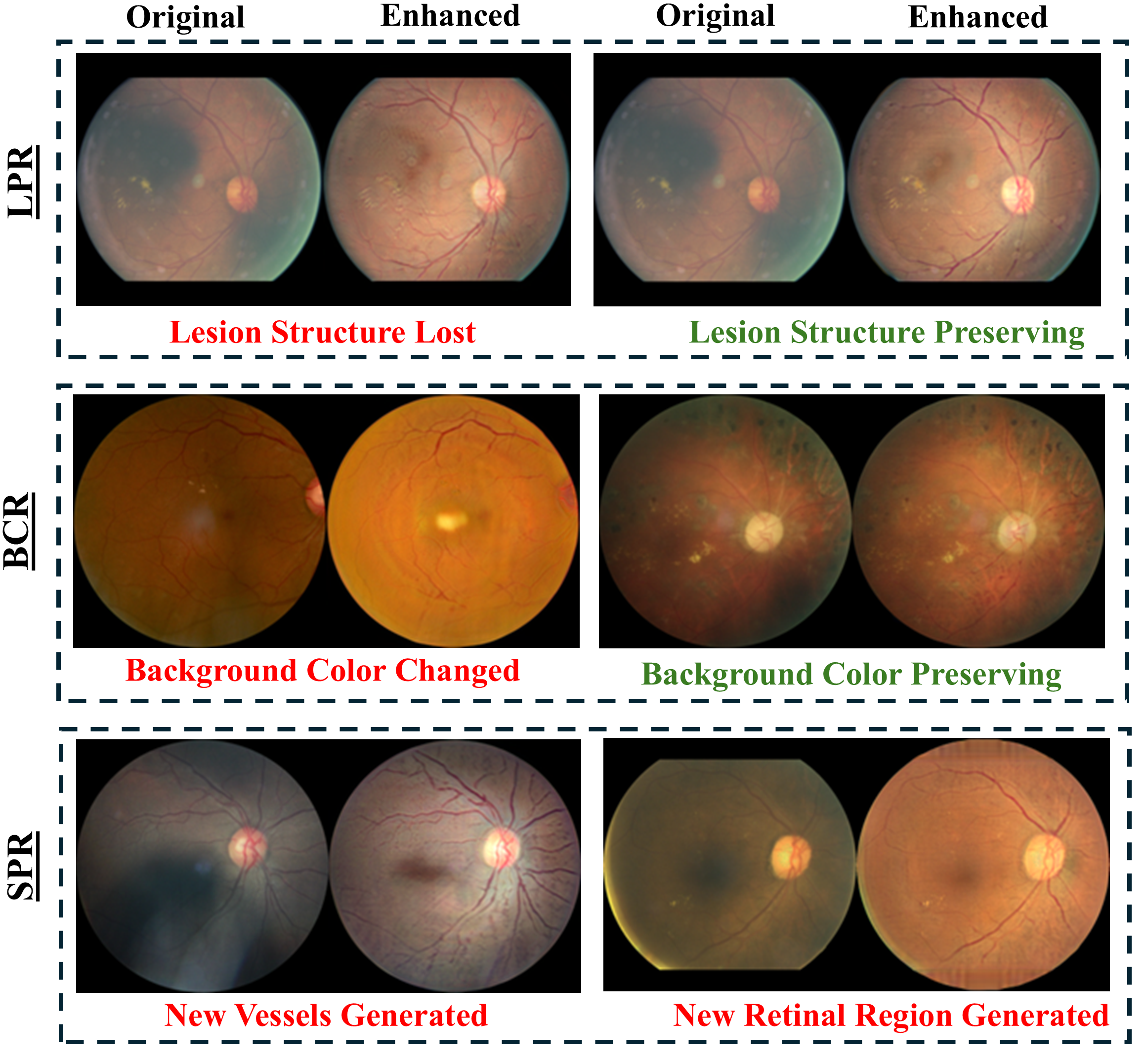}
    \caption{An illustrative medical expert clinical preference evaluation between (a) lesion preserving, (b) background preserving, and (c) structure-preserving.}
    \label{fig:expert-protocol}
\end{figure}
To better align with clinical preferences, we evaluated the enhanced images following protocols provided by medical experts. This evaluation includes the Background Preserving Ratio (BPR), Lesion Preserving Ratio (LPR), and Structure Preserving Ratio (SPR), each used to calculate the proportion of changes in the enhanced images. Importantly, we did not use all 2,434 testing images; instead, we selected 159 images with more prominent lesions, specifically those at DR grading levels 2, 3, and 4. These protocols are shown in Fig.~\ref{fig:expert-protocol}, which evaluate whether the denoised images maintain consistency with the original images regarding background, lesion, and structural integrity, helping to evaluate the practical applicability of these unpaired denoising models in real-world medical settings.

\begin{figure*}[t] \centering \includegraphics[width=\textwidth]{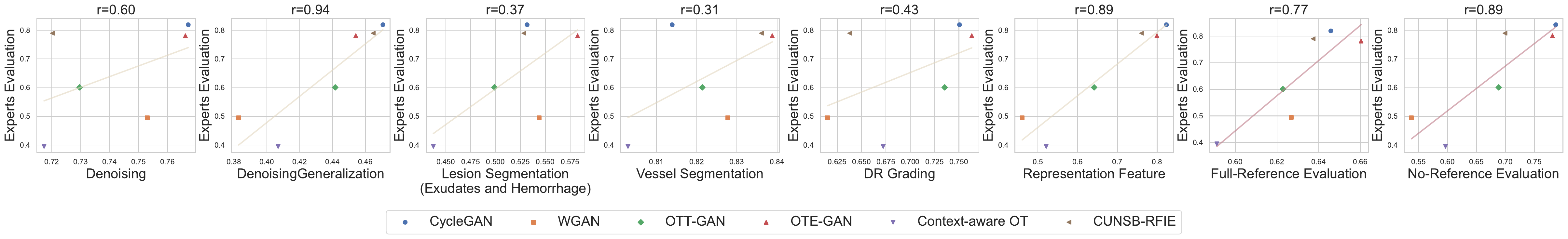}  
\caption{Validation of Expert Clinic Preference Alignment via Spearman’s correlation coefficient ($r$), which is used to assess the correlation between the Experts Protocol preference evaluation and other Eyebench evaluations. Single-dimension evaluations (e.g., denoising, segmentation) may show weak alignment with clinic preferences, while Eyebench multi-dimensional evaluations (e.g., Full-Reference, No-Reference) demonstrated stronger correlation.} 
\label{fig:correlation} 
\end{figure*}

\begin{figure*}[t] \centering \includegraphics[width=\textwidth]{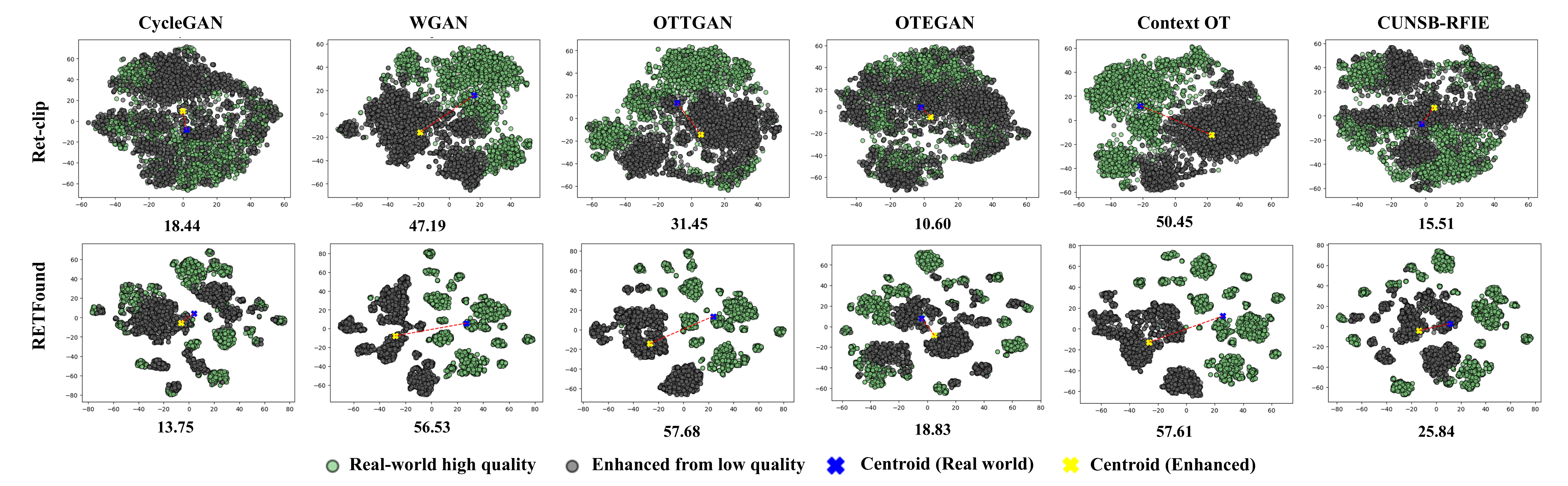}  
\caption{T-SNE visualizations of the latent representation features extracted from the RET-Clip and RETfound models. Closer proximity of the distributions indicates improved denoising performance of the unpaired method. This analysis demonstrates the effectiveness of the retrieval-enhanced frameworks in capturing and preserving meaningful feature representations. The Euclidean distance between the distribution centroids is showcased under each plot.} 
\label{fig:tsne} 
\end{figure*}


\subsection{Experiment Results}

\noindent\textbf{Full-Reference Evaluation.} Overall, paired methods outperform unpaired methods. As shown in Tab.~\ref{tb:deg-exp}, paired methods, particularly GFE-Net, effectively leverage frequency information, achieving higher SSIM (0.9554, 0.7935) and PSNR (29.719, 25.012) on EyeQ and IDRID, respectively. Among unpaired methods, CycleGAN and OTEGAN demonstrate competitive performance, especially on IDRID and DRIVE, where CycleGAN leads in SSIM (0.7668, 0.6681) and PSNR (22.511, 22.696), indicating robust noise reduction and generalization on unseen datasets. In segmentation tasks (Tab.~\ref{tab-seg}), SCR-Net achieves the highest AUC (0.9227), PR (0.7783), and F1 scores (0.7) in vessel segmentation among paired methods. Unpaired models CUNSB-RFIE and OTEGAN also perform comparably, with CUNSB-RFIE achieving the highest AUC (0.9163). For lesion segmentation, GFE-Net excels in HE lesions, while unpaired models CUNSB-RFIE and OTEGAN attain high F1 scores for EX lesions, demonstrating their effectiveness in lesion preservation.  

Since collecting paired noisy and clean images is challenging in real-world settings, unpaired methods are increasingly prioritized by medical experts.
Notably, some methods excel in noise reduction but face challenges in segmentation (e.g., CycleGAN), whereas SDE-based approaches like CUNSB-RFIE show strong generalization and excel in downstream segmentation. This highlights the need for multidimensional evaluation, as high noise reduction performance does not ensure the preservation of small, clinically significant structures.

\noindent\textbf{No-Reference Evaluation.}
Tab.~\ref{tab-noref} compares several methods for No-reference quality assessment in DR grading, Fréchet Inception Distance (FID) metrics, and Expert Protocol Evaluation. Each method's performance is evaluated across DR grading metrics (ACC, Kappa score, F1 score, AUC), FID scores (Retfound and Clip), and expert assessments (LPR, BPR, SPR). The best and second-best scores in each metric highlight the leading methods.
For DR grading, CycleGAN achieves the highest ACC (0.7588) and ranks second in Kappa, F1, and AUC, indicating strong grading capability. However, OTEGAN surpasses CycleGAN in overall quality metrics, with the highest Kappa (0.6433), F1 (0.7228), and AUC (0.9326), suggesting greater consistency and predictive accuracy for critical assessments. In FID metrics, which evaluate image realism and diversity, OTEGAN and CycleGAN excel. OTEGAN has the lowest FID-Clip score (11.114) and second-best FID-Retfound score (28.987), indicating superior image quality. CycleGAN scores best in FID-Retfound (23.778) and second-best in FID-Clip (11.530), showing strong but slightly less consistent image realism. Expert evaluations also favor OTEGAN and CycleGAN. CycleGAN achieves the highest SPR (0.8726), while OTEGAN and CUNSB-PRIE excels in LPR and BPR, with scores of 0.8280 and 0.8981 / 0.8535. these results suggest that the SDE-based method has more stable modality generation and clinic preference.

In summary, OTEGAN leads across multiple metrics, especially in DR grading, FID, and expert protocols, while CycleGAN follows closely, excelling in accuracy and realism. Other models, like Context-aware OT and CUNSB-RFIE, have strengths in specific areas but lack OTEGAN and CycleGAN consistency. This analysis underscores the effectiveness of OTEGAN in no-reference quality assessments, offering distinct advantages in image realism and expert evaluations.

\begin{figure}
     \centering
     \includegraphics[width=0.9\linewidth]{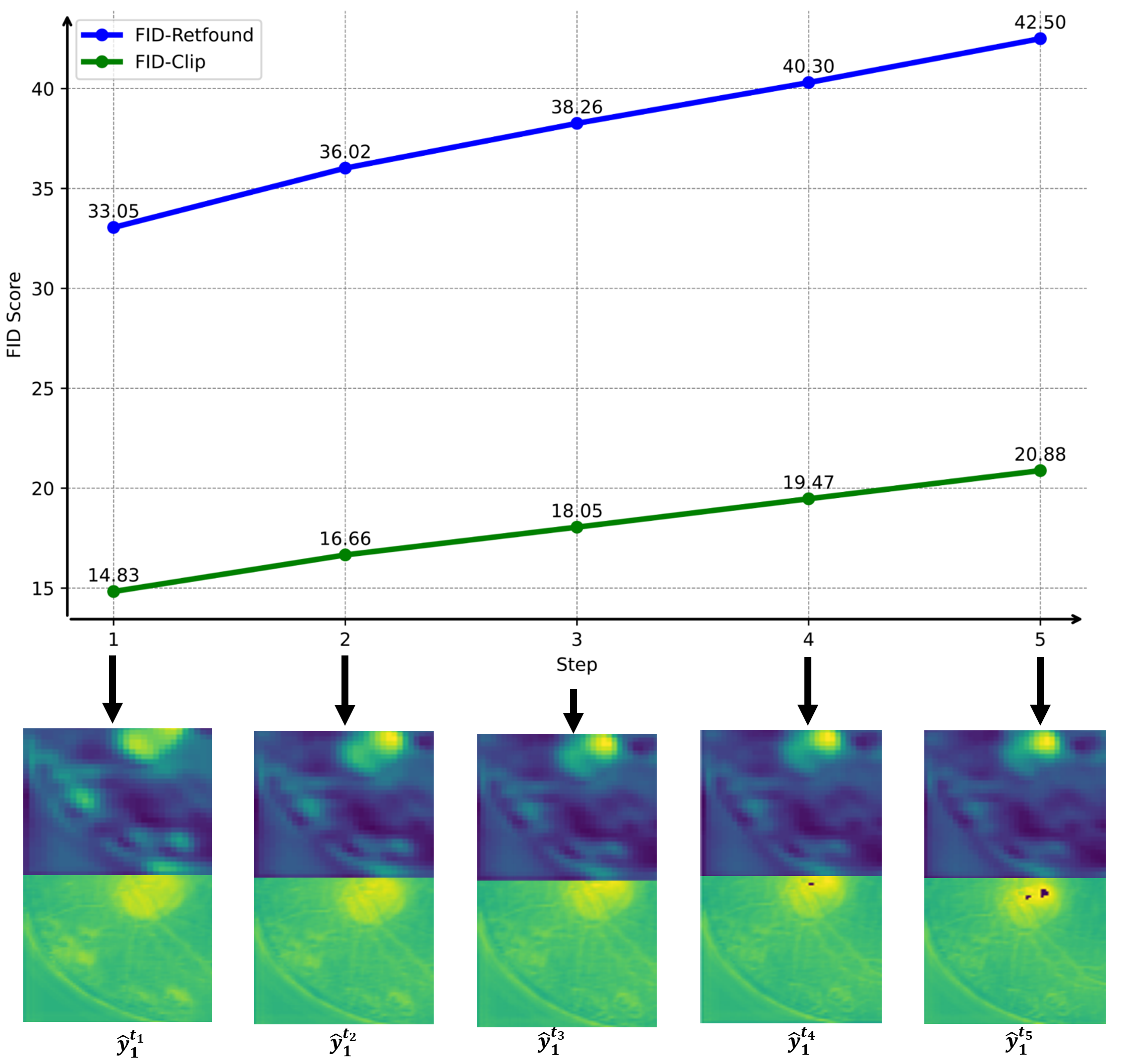}
     \caption{Illustration of denoising quality and skip connection feature patches of CUNSB-RFIE as steps $t_i$ increase. A higher FID score indicates lower quality, with the skip connection feature patches emphasizing lesion structures. This analysis demonstrates that high-frequency lesion details are gradually smoothed out over the denoising process.  }
     \label{fig:SB-noreference}
 \end{figure}

\section{Further Analysis}

\noindent\textbf{The necessity of multi-dimensional evaluation.} 
To further analyze the importance of the multi-dimensional evaluation of Eyebench, we visualized the correlation between medical experts guiding the protocol evaluation and other evaluations, including single-dimension and our multi-dimension evaluation (Full-Reference and No-Reference), as shown in Fig.~\ref{fig:correlation}. The results demonstrate that our multi-dimensional evaluation closely aligns with clinical preferences, whereas single-dimensional evaluations are likely to exhibit low correlation (e.g., denoising, DR grading, and segmentation). This finding also suggests that relying solely on a single task is insufficient to evaluate enhanced image quality. A multi-dimensional benchmark provides a more comprehensive comparison, offering medical experts greater insight and reference.

\noindent\textbf{Denoising Generalization Ability.} The evaluation of denoising methods necessitates a comparative analysis of the similarity between high-quality and enhanced-quality domains. We conducted a T-SNE~\cite{van2008visualizing} analysis to quantify this similarity and calculated the distance between centroids representing the real-world and enhanced domains, as illustrated in Fig.~\ref{fig:tsne}. Both OTEGAN and CycleGAN leverage high-quality prior learning, improving generalization capabilities. This high-quality regularization narrows the search space for GANs, enabling optimal transport between domains. Notably, the SDE-based method also shows strong potential to compete with GAN-based methods due to their stable modal-distribution modeling. In our Eyebench, CycleGAN, OTEGAN, and CUNSB-RFIE also demonstrated superior denoising generalization ability, indicating that high-quality regularization and SDE-based methods can enhance generalization capabilities.

\noindent\textbf{Trade off in GAN-based Methods.} We found that in these OT-based GANs, a regularization term is enforced between the noised and enhanced images to maintain consistency as enhanced images are transferred to a high-quality image domain. Since these two processes are trade-offs, contextual features or SSIM (structural similarity index) are often used to preserve lesions and vascular structures rather than noise. Thus, selecting an appropriate metric is crucial in this process. Depending on different downstream requirements, adjusting the weight of this regularization term is essential: too high a weight may prevent denoising, while too low a weight may cause the model to misclassify lesions and vessels as noise. A promising solution is the introduction of high-quality image priors in OTE-GAN and CycleGAN. CycleGAN employs cycle consistency, while OTE-GAN leverages structural consistency in high-quality images. This enables the model to partially learn structural and noise priors from high-quality images, preventing excessive structural alterations or incomplete noise modeling during denoising.

\noindent\textbf{Limitation in SDE-based method}. Since the SDE-based approach (i.e., CUNSB-RFIE) enforces smooth probabilistic distribution transport, we observed that high-frequency components, such as retinal lesions, were progressively smoothed out during the iterative generation process, resulting in a noticeable performance decline. We adopted the notation from~\cite{dong2024cunsb}, where $\hat{\mathbf{y}}_1^{t_i}$ for $i \in \{ 1, 2, 3, 4,5\}$ represents the progressively refined high-quality image counterparts. As shown in Fig.~\ref{fig:SB-noreference}, with each increment in step $i$, both FID-Retfound and FID-Clip scores increased, indicating a clear quality degradation. To further interpret this phenomenon, we visualized two skip connection features of the U-Net generator~\cite{dong2024cunsb}. As $t_i$ increases, the model progressively deactivates regions containing lesions. Due to the need for the SDE solver to generate $\hat{\mathbf{y}}_1^{t_i}$ to model the bridge $\mathbb{Q}^{SB}$ during training, high-frequency information tend to becomes smoothed out or diminished in the forward process, as Gaussian noise is added.

\section{Conclusion}
 
With the rapid development of generative models, aligning future methods for denoising fundus images with clinical needs has become essential. In this paper, we propose a new benchmark designed to provide more rigorous, clinically relevant evaluations of enhanced images, enabling broader access for medical experts. Furthermore, these multi-dimensional evaluations have demonstrated a strong correlation with manual expert evaluations, helping to bridge the gap in applying generative model-based denoising methods to real-world clinical requirements.

\noindent\textbf{Limitations and Future Work.} Currently, our evaluations are primarily based on deep learning methods. In the future, we plan to expand our work to include more unsupervised traditional algorithms and apply these methods to MRI enhancement tasks.

\noindent\textbf{Ethics Statement.} This retrospective study used open-access human subject data and did not require ethical approval, as confirmed by their license~\cite{fu2019evaluation,diabetic-retinopathy-detection,idrid,drive}.

{
    \small
    \bibliographystyle{ieeenat_fullname}
    \bibliography{main}

\begin{thebibliography}{41}
\providecommand{\natexlab}[1]{#1}
\providecommand{\url}[1]{\texttt{#1}}
\expandafter\ifx\csname urlstyle\endcsname\relax
  \providecommand{\doi}[1]{doi: #1}\else
  \providecommand{\doi}{doi: \begingroup \urlstyle{rm}\Url}\fi

\bibitem[Arjovsky et~al.(2017)Arjovsky, Chintala, and Bottou]{arjovsky2017wasserstein}
Martin Arjovsky, Soumith Chintala, and L{\'e}on Bottou.
\newblock Wasserstein generative adversarial networks.
\newblock In \emph{International conference on machine learning}, pages 214--223. PMLR, 2017.

\bibitem[Brunet et~al.(2011)Brunet, Vrscay, and Wang]{brunet2011mathematical}
Dominique Brunet, Edward~R Vrscay, and Zhou Wang.
\newblock On the mathematical properties of the structural similarity index.
\newblock \emph{IEEE Transactions on Image Processing}, 21\penalty0 (4):\penalty0 1488--1499, 2011.

\bibitem[Cheng et~al.(2021)Cheng, Lin, Huang, Lyu, and Tang]{i-secret}
Pujin Cheng, Li Lin, Yijin Huang, Junyan Lyu, and Xiaoying Tang.
\newblock I-secret: Importance-guided fundus image enhancement via semi-supervised contrastive constraining.
\newblock In \emph{Medical Image Computing and Computer Assisted Intervention--MICCAI 2021: 24th International Conference, Strasbourg, France, September 27--October 1, 2021, Proceedings, Part VIII 24}, pages 87--96. Springer, 2021.

\bibitem[Deng et~al.(2022)Deng, Cai, Chen, Gong, Bao, Yao, Fang, Yang, Zhang, and Ma]{deng2022rformer}
Zhuo Deng, Yuanhao Cai, Lu Chen, Zheng Gong, Qiqi Bao, Xue Yao, Dong Fang, Wenming Yang, Shaochong Zhang, and Lan Ma.
\newblock Rformer: Transformer-based generative adversarial network for real fundus image restoration on a new clinical benchmark.
\newblock \emph{IEEE Journal of Biomedical and Health Informatics}, 26\penalty0 (9):\penalty0 4645--4655, 2022.

\bibitem[Dong et~al.(2024)Dong, Vasa, Zhu, Qiu, Chen, Su, Xiong, Yang, Chen, and Wang]{dong2024cunsb}
Xuanzhao Dong, Vamsi~Krishna Vasa, Wenhui Zhu, Peijie Qiu, Xiwen Chen, Yi Su, Yujian Xiong, Zhangsihao Yang, Yanxi Chen, and Yalin Wang.
\newblock Cunsb-rfie: Context-aware unpaired neural schr"$\{$o$\}$ dinger bridge in retinal fundus image enhancement.
\newblock \emph{arXiv preprint arXiv:2409.10966}, 2024.

\bibitem[Du et~al.(2024)Du, Guo, Zhang, Yang, Liu, Li, and Wang]{du2024ret}
Jiawei Du, Jia Guo, Weihang Zhang, Shengzhu Yang, Hanruo Liu, Huiqi Li, and Ningli Wang.
\newblock Ret-clip: A retinal image foundation model pre-trained with clinical diagnostic reports.
\newblock \emph{arXiv preprint arXiv:2405.14137}, 2024.

\bibitem[Dugas et~al.(2015)Dugas, Jared, Jorge, and Cukierski]{diabetic-retinopathy-detection}
Emma Dugas, Jared, Jorge, and Will Cukierski.
\newblock Diabetic retinopathy detection.
\newblock \url{https://kaggle.com/competitions/diabetic-retinopathy-detection}, 2015.
\newblock Kaggle.

\bibitem[Dumitrascu et~al.(2022)Dumitrascu, Zhu, Qiu, Nandakumar, and Wang]{deeplearning3}
Oana~M Dumitrascu, Wenhui Zhu, Peijie Qiu, Keshav Nandakumar, and Yalin Wang.
\newblock Automated retinal imaging analysis for alzheimers disease screening.
\newblock In \emph{IEEE International Symposium on Biomedical Imaging: From Nano to Macro (ISBI)}, 2022.

\bibitem[Fu et~al.(2019)Fu, Wang, Shen, Cui, Xu, Liu, and Shao]{fu2019evaluation}
Huazhu Fu, Boyang Wang, Jianbing Shen, Shanshan Cui, Yanwu Xu, Jiang Liu, and Ling Shao.
\newblock Evaluation of retinal image quality assessment networks in different color-spaces.
\newblock In \emph{Medical Image Computing and Computer Assisted Intervention--MICCAI 2019: 22nd International Conference, Shenzhen, China, October 13--17, 2019, Proceedings, Part I 22}, pages 48--56. Springer, 2019.

\bibitem[Goodfellow et~al.(2020)Goodfellow, Pouget-Abadie, Mirza, Xu, Warde-Farley, Ozair, Courville, and Bengio]{goodfellow2020generative}
Ian Goodfellow, Jean Pouget-Abadie, Mehdi Mirza, Bing Xu, David Warde-Farley, Sherjil Ozair, Aaron Courville, and Yoshua Bengio.
\newblock Generative adversarial networks.
\newblock \emph{Communications of the ACM}, 63\penalty0 (11):\penalty0 139--144, 2020.

\bibitem[Gulrajani et~al.(2017)Gulrajani, Ahmed, Arjovsky, Dumoulin, and Courville]{gulrajani2017improved}
Ishaan Gulrajani, Faruk Ahmed, Martin Arjovsky, Vincent Dumoulin, and Aaron~C Courville.
\newblock Improved training of wasserstein gans.
\newblock \emph{Advances in neural information processing systems}, 30, 2017.

\bibitem[Ho et~al.(2020)Ho, Jain, and Abbeel]{ho2020denoising}
Jonathan Ho, Ajay Jain, and Pieter Abbeel.
\newblock Denoising diffusion probabilistic models.
\newblock \emph{Advances in neural information processing systems}, 33:\penalty0 6840--6851, 2020.

\bibitem[Isola et~al.(2017)Isola, Zhu, Zhou, and Efros]{isola2017image}
Phillip Isola, Jun-Yan Zhu, Tinghui Zhou, and Alexei~A Efros.
\newblock Image-to-image translation with conditional adversarial networks.
\newblock In \emph{Proceedings of the IEEE conference on computer vision and pattern recognition}, pages 1125--1134, 2017.

\bibitem[Li et~al.(2022)Li, Liu, Fu, Shu, Zhao, Luo, Hu, and Liu]{li2022structure}
Heng Li, Haofeng Liu, Huazhu Fu, Hai Shu, Yitian Zhao, Xiaoling Luo, Yan Hu, and Jiang Liu.
\newblock Structure-consistent restoration network for cataract fundus image enhancement.
\newblock In \emph{International Conference on Medical Image Computing and Computer-Assisted Intervention}, pages 487--496. Springer, 2022.

\bibitem[Li et~al.(2023)Li, Liu, Fu, Xu, Shu, Niu, Hu, and Liu]{li2023generic}
Heng Li, Haofeng Liu, Huazhu Fu, Yanwu Xu, Hai Shu, Ke Niu, Yan Hu, and Jiang Liu.
\newblock A generic fundus image enhancement network boosted by frequency self-supervised representation learning.
\newblock \emph{Medical Image Analysis}, 90:\penalty0 102945, 2023.

\bibitem[Liu et~al.(2022)Liu, Li, Fu, Xiao, Gao, Hu, and Liu]{10.1007/978-3-031-16434-7_49}
Haofeng Liu, Heng Li, Huazhu Fu, Ruoxiu Xiao, Yunshu Gao, Yan Hu, and Jiang Liu.
\newblock Degradation-invariant enhancement of fundus images via pyramid constraint network.
\newblock In \emph{Medical Image Computing and Computer Assisted Intervention -- MICCAI 2022}, pages 507--516, Cham, 2022. Springer Nature Switzerland.

\bibitem[Mao et~al.(2017)Mao, Li, Xie, Lau, Wang, and Paul~Smolley]{mao2017least}
Xudong Mao, Qing Li, Haoran Xie, Raymond~YK Lau, Zhen Wang, and Stephen Paul~Smolley.
\newblock Least squares generative adversarial networks.
\newblock In \emph{Proceedings of the IEEE international conference on computer vision}, pages 2794--2802, 2017.

\bibitem[Mechrez et~al.(2018)Mechrez, Talmi, and Zelnik-Manor]{mechrez2018contextual}
Roey Mechrez, Itamar Talmi, and Lihi Zelnik-Manor.
\newblock The contextual loss for image transformation with non-aligned data, 2018.

\bibitem[Park et~al.(2020)Park, Efros, Zhang, and Zhu]{park2020contrastive}
Taesung Park, Alexei~A Efros, Richard Zhang, and Jun-Yan Zhu.
\newblock Contrastive learning for unpaired image-to-image translation.
\newblock In \emph{Computer Vision--ECCV 2020: 16th European Conference, Glasgow, UK, August 23--28, 2020, Proceedings, Part IX 16}, pages 319--345. Springer, 2020.

\bibitem[Porwal and et~al.(2018)]{idrid}
Prasanna Porwal and et al.
\newblock Idrid: A database for diabetic retinopathy screening research.
\newblock \emph{Data}, 3\penalty0 (3), 2018.

\bibitem[Qian et~al.(2024)Qian, Sheng, Chen, Wang, Li, Jin, Guan, Jiang, Wu, Wang, et~al.]{deeplearning5}
Bo Qian, Bin Sheng, Hao Chen, Xiangning Wang, Tingyao Li, Yixiao Jin, Zhouyu Guan, Zehua Jiang, Yilan Wu, Jinyuan Wang, et~al.
\newblock A competition for the diagnosis of myopic maculopathy by artificial intelligence algorithms.
\newblock \emph{JAMA ophthalmology}, 2024.

\bibitem[Ronneberger et~al.(2015)Ronneberger, Fischer, and Brox]{ronneberger2015unet}
Olaf Ronneberger, Philipp Fischer, and Thomas Brox.
\newblock U-net: Convolutional networks for biomedical image segmentation, 2015.

\bibitem[Salmona et~al.(2022)Salmona, De~Bortoli, Delon, and Desolneux]{salmona2022can}
Antoine Salmona, Valentin De~Bortoli, Julie Delon, and Agnes Desolneux.
\newblock Can push-forward generative models fit multimodal distributions?
\newblock \emph{Advances in Neural Information Processing Systems}, 35:\penalty0 10766--10779, 2022.

\bibitem[Shen et~al.(2020)Shen, Fu, Shen, and Shao]{shen2020modeling}
Ziyi Shen, Huazhu Fu, Jianbing Shen, and Ling Shao.
\newblock Modeling and enhancing low-quality retinal fundus images.
\newblock \emph{IEEE transactions on medical imaging}, 40\penalty0 (3):\penalty0 996--1006, 2020.

\bibitem[Song et~al.(2020)Song, Meng, and Ermon]{song2020denoising}
Jiaming Song, Chenlin Meng, and Stefano Ermon.
\newblock Denoising diffusion implicit models.
\newblock \emph{arXiv preprint arXiv:2010.02502}, 2020.

\bibitem[Song and Ermon(2019)]{song2019generative}
Yang Song and Stefano Ermon.
\newblock Generative modeling by estimating gradients of the data distribution.
\newblock \emph{Advances in neural information processing systems}, 32, 2019.

\bibitem[Staal and et~al.(2004)]{drive}
J. Staal and et al.
\newblock {{R}idge-based vessel segmentation in color images of the retina}.
\newblock \emph{IEEE Trans Med Imaging}, 23\penalty0 (4):\penalty0 501--509, 2004.

\bibitem[Tong et~al.(2023)Tong, Malkin, Huguet, Zhang, Rector-Brooks, Fatras, Wolf, and Bengio]{tong2023improving}
Alexander Tong, Nikolay Malkin, Guillaume Huguet, Yanlei Zhang, Jarrid Rector-Brooks, Kilian Fatras, Guy Wolf, and Yoshua Bengio.
\newblock Improving and generalizing flow-based generative models with minibatch optimal transport.
\newblock \emph{arXiv preprint arXiv:2302.00482}, 2023.

\bibitem[Van~der Maaten and Hinton(2008)]{van2008visualizing}
Laurens Van~der Maaten and Geoffrey Hinton.
\newblock Visualizing data using t-sne.
\newblock \emph{Journal of machine learning research}, 9\penalty0 (11), 2008.

\bibitem[Vasa et~al.(2024)Vasa, Qiu, Zhu, Xiong, Dumitrascu, and Wang]{vasa2024context}
Vamsi~Krishna Vasa, Peijie Qiu, Wenhui Zhu, Yujian Xiong, Oana Dumitrascu, and Yalin Wang.
\newblock Context-aware optimal transport learning for retinal fundus image enhancement.
\newblock \emph{arXiv preprint arXiv:2409.07862}, 2024.

\bibitem[Villani et~al.(2009)]{villani2009optimal}
C{\'e}dric Villani et~al.
\newblock \emph{Optimal transport: old and new}.
\newblock Springer, 2009.

\bibitem[Wang et~al.()Wang, Zhu, Qin, Li, Dumitrascu, Chen, Qiu, Razi, and Wang]{deeplearning4}
Hao Wang, Wenhui Zhu, Jiayou Qin, Xin Li, Oana Dumitrascu, Xiwen Chen, Peijie Qiu, Abolfazl Razi, and Yalin Wang.
\newblock Rbad: A dataset and benchmark for retinal bifurcation angle detection.
\newblock In \emph{IEEE-EMBS International Conference on Biomedical and Health Informatics}.

\bibitem[Wang et~al.(2022)Wang, Wen, Yan, and Liu]{wang2022optimal}
Wei Wang, Fei Wen, Zeyu Yan, and Peilin Liu.
\newblock Optimal transport for unsupervised denoising learning.
\newblock \emph{IEEE Transactions on Pattern Analysis and Machine Intelligence}, 45\penalty0 (2):\penalty0 2104--2118, 2022.

\bibitem[Wang et~al.(2003)Wang, Simoncelli, and Bovik]{wang2003multiscale}
Zhou Wang, Eero~P Simoncelli, and Alan~C Bovik.
\newblock Multiscale structural similarity for image quality assessment.
\newblock In \emph{The Thrity-Seventh Asilomar Conference on Signals, Systems \& Computers, 2003}, pages 1398--1402. Ieee, 2003.

\bibitem[Zhou et~al.(2023)Zhou, Chia, Wagner, Ayhan, Williamson, Struyven, Liu, Xu, Lozano, Woodward-Court, et~al.]{zhou2023foundation}
Yukun Zhou, Mark~A Chia, Siegfried~K Wagner, Murat~S Ayhan, Dominic~J Williamson, Robbert~R Struyven, Timing Liu, Moucheng Xu, Mateo~G Lozano, Peter Woodward-Court, et~al.
\newblock A foundation model for generalizable disease detection from retinal images.
\newblock \emph{Nature}, 622\penalty0 (7981):\penalty0 156--163, 2023.

\bibitem[Zhu et~al.(2017)Zhu, Park, Isola, and Efros]{cyclegan}
Jun{-}Yan Zhu, Taesung Park, Phillip Isola, and Alexei~A. Efros.
\newblock {Unpaired Image-to-Image Translation Using Cycle-Consistent Adversarial Networks}.
\newblock \emph{{CVPR}}, pages 2242--2251, 2017.

\bibitem[Zhu et~al.(2023{\natexlab{a}})Zhu, Qiu, Chen, Li, Wang, Lepore, Dumitrascu, and Wang]{deeplearning6}
Wenhui Zhu, Peijie Qiu, Xiwen Chen, Huayu Li, Hao Wang, Natasha Lepore, Oana~M Dumitrascu, and Yalin Wang.
\newblock Beyond mobilenet: An improved mobilenet for retinal diseases.
\newblock In \emph{International Conference on Medical Image Computing and Computer-Assisted Intervention}, pages 56--65. Springer, 2023{\natexlab{a}}.

\bibitem[Zhu et~al.(2023{\natexlab{b}})Zhu, Qiu, Dumitrascu, Sobczak, Farazi, Yang, Nandakumar, and Wang]{zhu2023otre}
Wenhui Zhu, Peijie Qiu, Oana~M Dumitrascu, Jacob~M Sobczak, Mohammad Farazi, Zhangsihao Yang, Keshav Nandakumar, and Yalin Wang.
\newblock Otre: Where optimal transport guided unpaired image-to-image translation meets regularization by enhancing.
\newblock In \emph{International Conference on Information Processing in Medical Imaging}, pages 415--427. Springer, 2023{\natexlab{b}}.

\bibitem[Zhu et~al.(2023{\natexlab{c}})Zhu, Qiu, Farazi, Nandakumar, Dumitrascu, and Wang]{zhu2023optimal}
Wenhui Zhu, Peijie Qiu, Mohammad Farazi, Keshav Nandakumar, Oana~M Dumitrascu, and Yalin Wang.
\newblock Optimal transport guided unsupervised learning for enhancing low-quality retinal images.
\newblock \emph{Proc IEEE Int Symp Biomed Imaging}, 2023{\natexlab{c}}.

\bibitem[Zhu et~al.(2023{\natexlab{d}})Zhu, Qiu, Lepore, Dumitrascu, and Wang]{deeplearning2}
Wenhui Zhu, Peijie Qiu, Natasha Lepore, Oana~M Dumitrascu, and Yalin Wang.
\newblock Self-supervised equivariant regularization reconciles multiple-instance learning: Joint referable diabetic retinopathy classification and lesion segmentation.
\newblock In \emph{18th International Symposium on Medical Information Processing and Analysis}, pages 100--107. SPIE, 2023{\natexlab{d}}.

\bibitem[Zhu et~al.(2024)Zhu, Qiu, Chen, Li, Lepore, Dumitrascu, and Wang]{deeplearning1}
Wenhui Zhu, Peijie Qiu, Xiwen Chen, Xin Li, Natasha Lepore, Oana~M. Dumitrascu, and Yalin Wang.
\newblock nnmobilenet: Rethinking cnn for retinopathy research.
\newblock In \emph{Proceedings of the IEEE/CVF Conference on Computer Vision and Pattern Recognition (CVPR) Workshops}, pages 2285--2294, 2024.

\end{thebibliography}
}

 \clearpage\section*{\Large Supplementary Materials - DGR-MIL: Exploring Diverse Global Representation in Multiple Instance Learning for Whole Slide Image Classification}

\thispagestyle{empty}
\appendix

\section{Full-Reference Quality Assessment Experiments Details }\label{Sec:full-reference}

\subsection{Datasets.} For full-reference assessment, we used the previously synthesized Full-Reference Evaluation Dataset. We strictly followed the training configurations for paired and unpaired methods. For the unpaired method, synthetic low-quality images from the training set $A$ (i.e., $A^{\ast}$) were used as input images, while high-quality images from the training set $B$ served as the clean reference images. For the paired method, we performed supervised training using low-high-quality image pairs from the training set $A$ (i.e., $A^{\ast}$ and $A$).

\subsection{SCR-Net~\cite{li2022structure}}
The model was trained for 150 epochs using Adam optimizer, with an initial learning rate of $2 \times 10^{-4}$ and $\beta_1$ value set to $0.5$, followed by 50 epochs with a learning rate linearly decayed to $0$. The training batch size was 32. All images were resized to $ 256 \times 256$ with a random flipping data augmentation technique. For model architectures, the generator and discriminator architectures followed the architectures and configurations described in~\cite{li2022structure}.



\subsection{Cofe-Net~\cite{shen2020modeling}}

The model was trained for 300 epochs using the SGD optimizer, with an initial learning rate of $1 \times 10^{-4}$, which was gradually reduced to 0 over the final 150 epochs. The training batch size was 16, and all images were resize to $512 \times 512$.

The loss function comprised four components: main scale error loss ($L_m$), multiple-scale pixel loss ($L^s_p$), multiple-scale content loss ($L^s_c$) and RSA module loss ($L_v$), as described in~\cite{shen2020modeling}, where the $s$ denotes the scale index. The weight for $L^s_p$, $L^s_c$ and $L_v$ was set to $\lambda_p=10$, $\lambda_c=1$ and $\lambda_v=0.1$, respectively, during the training process.

\subsection{PCE-Net~\cite{10.1007/978-3-031-16434-7_49}}

The model was trained for 200 epochs using the Adam optimizer, with an initial learning rate of $1 \times 10^{-3}$, which was gradually reduced to 0 over the final 50 epochs. The training batch size was 4, and all input images were resized to $256 \times 256$. Data augmentation strategies, including random horizontal and vertical flips with a probability of 0.5, were applied to enhance generalization.

The loss function comprised two components: enhancement loss ($L_E$) and the weighted feature pyramid constraint loss ($L_C$), as described in~\cite{10.1007/978-3-031-16434-7_49}. The weight for $L_C$ was set to $\lambda_C=0.1$ during the training process. Additionally, we adopted a U-Net architecture proposed in ~\cite{10.1007/978-3-031-16434-7_49}.

\subsection{GFE-Net~\cite{li2023generic}}

The model was trained for 200 epochs using the Adam optimizer, with an initial learning rate of $1\times 10^{-3}$, which was gradually reduced to 0 over the final 50 epochs. The training batch size was set to 4, and all input images were resized to $256 \times 256$. Data augmentation strategies, including random horizontal and vertical flips with a probability of 0.5, were applied to enhance generalization.

We employed the same weight (e.g., $\lambda_{all}$ = 1) for all loss losses, including enhancement loss, cycle-consistency loss, and reconstruction loss. Furthermore, we adopted the architecture proposed in~\cite{li2023generic}, implementing a symmetric U-Net with 8 down-sampling and 8 up-sampling layers.

\subsection{I-SECRET~\cite{i-secret}}

The model was trained for 200 epochs using Adam optimizer with an initial learning rate of $1 \times 10^{-4}$ and $\beta$ values set to $0.5$ and $0.999$, respectively. The learning rate followed a cosine decay schedule. The training batch size was set to 8. All images were resized to $256 \times 256$ with random cropping and flipping augmentation strategies.

For model architectures, the generator consisted of 2 down-sampling layers, each with 64 filters and 9 residual blocks. Input and output channels were set to 3 for RGB inputs. The discriminator included 64 filters and 3 layers. Instance normalization and reflective padding were used. The training process employed a least-squares GAN loss~\cite{mao2017least}, a ResNet-based generator, and a PatchGAN-based~\cite{isola2017image} discriminator. GAN and reconstruction losses were weighted at $1.0$, while their importance with the contrastive loss (ICC-loss) and importance-guided supervised loss (IS-loss)~\cite{i-secret} were enabled with weights of $1.0$.

\subsection{ RFormer~\cite{deng2022rformer}.}
The model was trained for 150 epochs using Adam optimizer, with an initial learning rate of $1 \times 10^{-4}$ and $\beta$ values set to 0.9 and 0.999, respectively. The cosine annealing strategy was employed to steadily decrease the learning rate from the initial value to $1 \times 10^{-6}$ during the training procedure. The training batch size was set to 32. All images were resized to $256 \times 256$ without any additional augmentation strategies. The model architecture followed the design proposed in~\cite{deng2022rformer}, which was consistently maintained throughout our experiments.


\subsection{CycleGAN~\cite{cyclegan}, WGAN~\cite{gulrajani2017improved}, OTTGAN~\cite{wang2022optimal}, OTEGAN~\cite{zhu2023optimal}  }
The models were trained for 200 epochs using the RMSprop optimizer, with initial learning rates for the generator and discriminator set to $0.5 \times 10^{-4}$ and $1 \times 10^{-4}$, respectively. The learning rate followed a linear decay schedule, decreasing by a factor of 10 every 100 epochs. The training batch size was set to 2. All input images were resized to $256 \times 256$, with random horizontal and vertical flips applied as augmentation strategies.
For CycleGAN, the weighting parameters in the final objective were set to $\lambda_{GAN} = 1$, $\lambda_{Cycle} = 10$, and $\lambda_{Idt} = 5$, corresponding to the weights for the GAN loss, cycle consistency loss, and identity loss, respectively. The Mean Squared Error (MSE) loss was used for the GAN loss, while the cycle consistency and identity losses were computed using the L1-norm. For OTTGAN and OTEGAN, the weighting parameter $\lambda_{OT}$ was set to 40, representing the optimal transport (OT) cost. Furthermore, the OT loss was calculated using the MSE loss for OTTGAN and the MS-SSIM loss for OTEGAN. The generator and discriminator architectures were implemented following the baseline designs described in~\cite{zhu2023optimal,zhu2023otre}.

\subsection{Context-aware OT~\cite{vasa2024context}}
The model was trained for 200 epochs using the RMSprop optimizer, with initial learning rates for the generator and discriminator set to $0.5 \times 10^{-4}$ and $1 \times 10^{-4}$, respectively. The learning rate followed a linear decay schedule, decreasing by a factor of 10 every 50 epochs. The training batch size was set to 2. All input images were resized to $256 \times 256$ without additional augmentation strategies.
A warm-up training strategy was employed, wherein the context-OT loss was introduced after the first 50 epochs. The weighting parameter for this loss was set to $5\times10^{-2}$. We utilized a pre-trained VGG~\cite{mechrez2018contextual} network outlined in~\cite{vasa2024context} to compute the OT loss at feature spaces.
The generator and discriminator architectures followed the designs outlined in~\cite{vasa2024context}.

\subsection{CUNSB-RFIE~\cite{dong2024cunsb}}
The model was trained for 130 epochs using the Adam optimizer, with an initial learning rate of $2 \times 10^ {-4}$. The learning rate was linearly decayed to 0 after the first 80 epochs, and the batch size was set to 8. All input images were resized to $256 \times 256$ without applying any additional augmentation strategies.

The weighting parameters in the final objective were set as $\lambda_{SB} = 1$, $\lambda_{SSIM} = 0.8$, and $\lambda_{NCE} = 1$, corresponding to the weights for entropy-regularized OT loss, task-specific regularization with MS-SSIM~\cite{brunet2011mathematical}, and PatchNCE~\cite{park2020contrastive} loss, respectively.

The generator and discriminator architectures followed the designs described in~\cite{dong2024cunsb}. Specifically, the base number of channels for the generator was set to 32, and 9 ResNet blocks were used in the bottleneck. In addition to the output features of all downsampling layers, the bottleneck's input and middle feature maps were utilized to calculate the PatchNCE regularization.

\subsection{Vessel Segmentation}
A vanilla U-Net model~\cite{ronneberger2015unet} was employed for the downstream vessel segmentation task. The network comprised 4 layers with a base channel size 64 and a channel scale expansion ratio of 2. The training was conducted over 10 epochs using the Adam optimizer, with a batch size of 64 and an initial learning rate of $5 \times 10^{-5}$, which followed a cosine annealing learning rate scheduler. 

Before training, the enhanced images and their corresponding ground-truth vessel segmentation masks were preprocessed. The preprocessing pipeline included random cropping to $ 48 \times 48$ patches, followed by data augmentation techniques such as random horizontal flips, random vertical flips (with a probability of 0.5), and random rotation.

\begin{figure}[ht]
    \centering
    \includegraphics[width=1\linewidth]{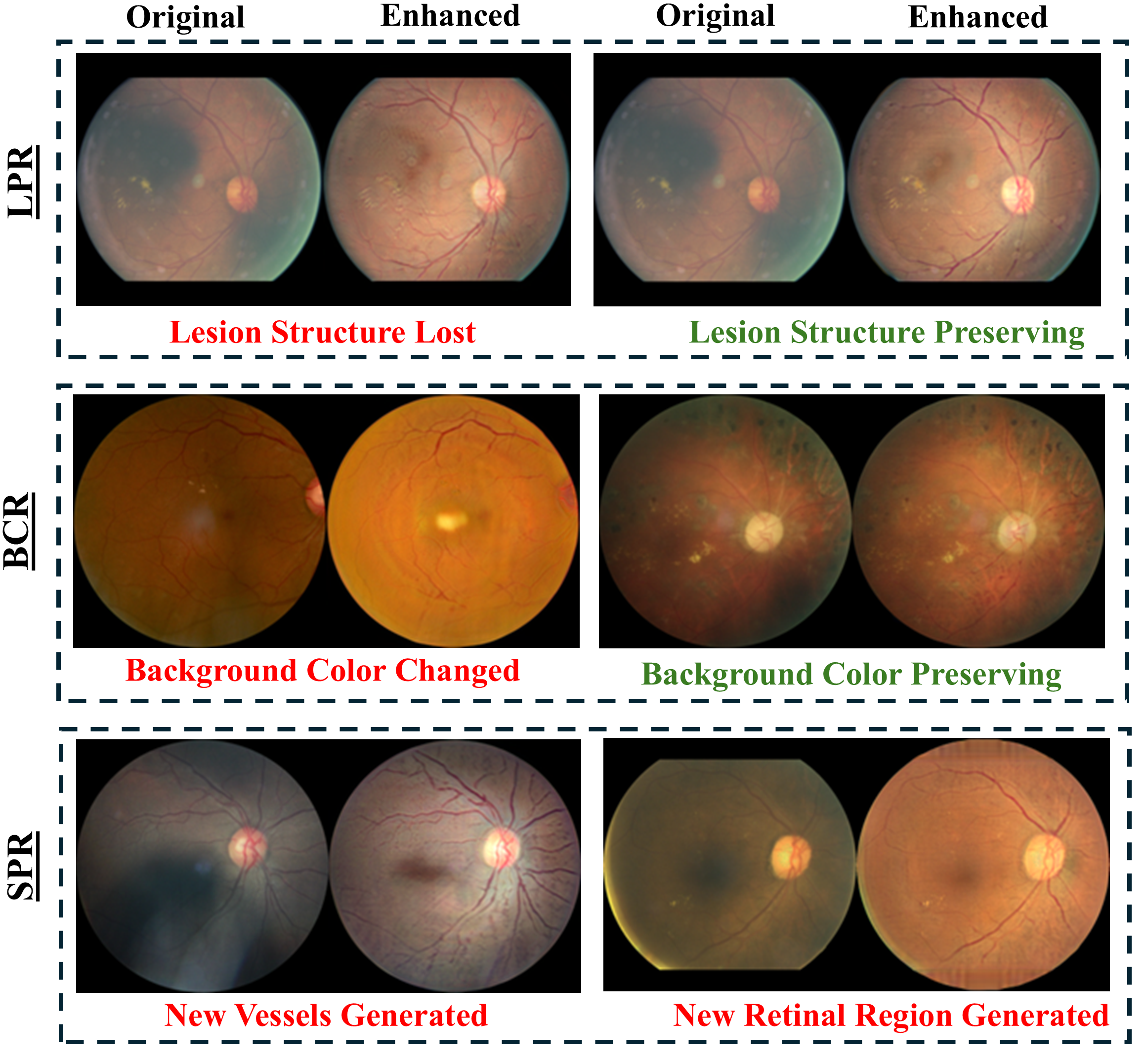}
    \caption{An illustrative medical expert clinical preference evaluation between (a) lesion preserving, (b) background preserving, and (c) structure-preserving.}
    \label{fig:expert-protocol}
\end{figure}

\begin{figure*}[t]
  \centering
  \includegraphics[width=\textwidth]{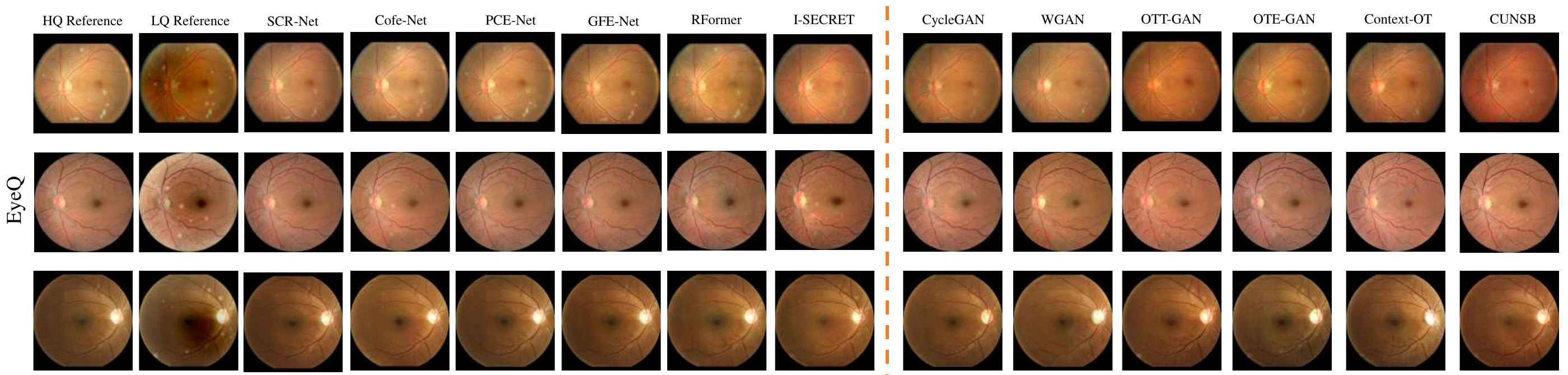}  
  \caption{ Illustration of the Denoising Evaluation on the EyeQ dataset. The first and second columns show the high- and low-quality image references, respectively, while the remaining columns display the synthetic high-quality images generated by all baseline models.
  }
  \label{fig:full-reference-eyeq}
\end{figure*}
\begin{figure*}[t]
  \centering
  \includegraphics[width=\textwidth]{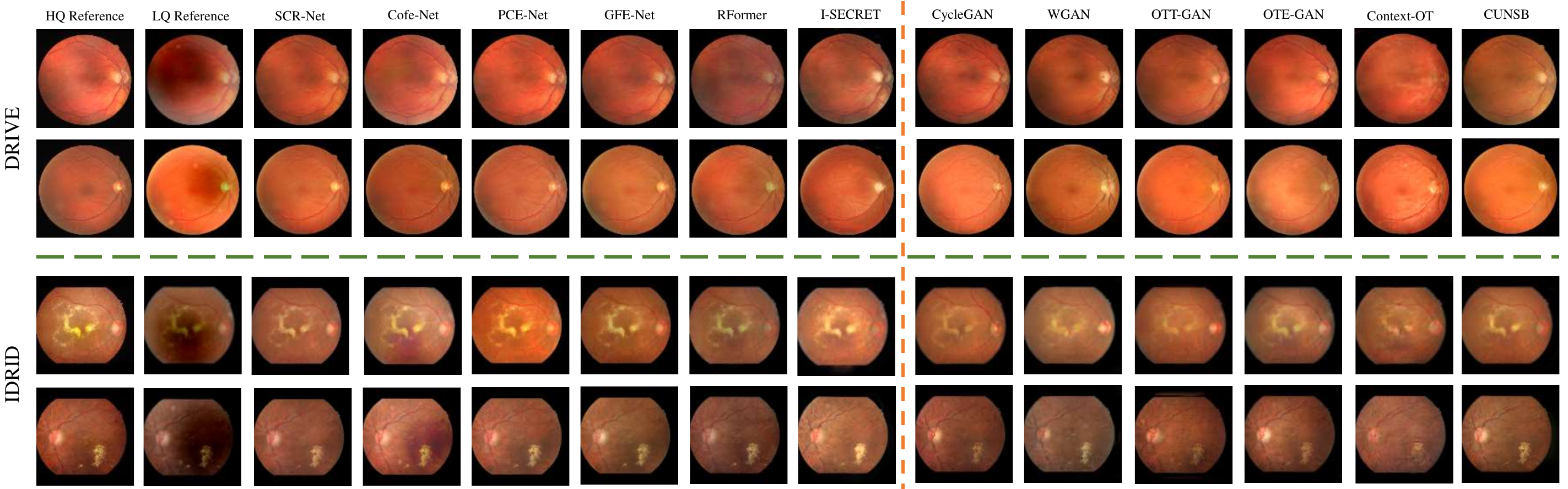}  
  \caption{ Illustration of the Denoising Generalization Evaluation on the DRIVE and IDRID datasets. The first and second columns show the high- and low-quality image references, respectively, while the remaining columns display the synthetic high-quality images generated by all baseline models.
  }
  \label{fig:full-reference-generalization}
\end{figure*}
\begin{figure*}[t]
  \centering
  \includegraphics[width=\textwidth]{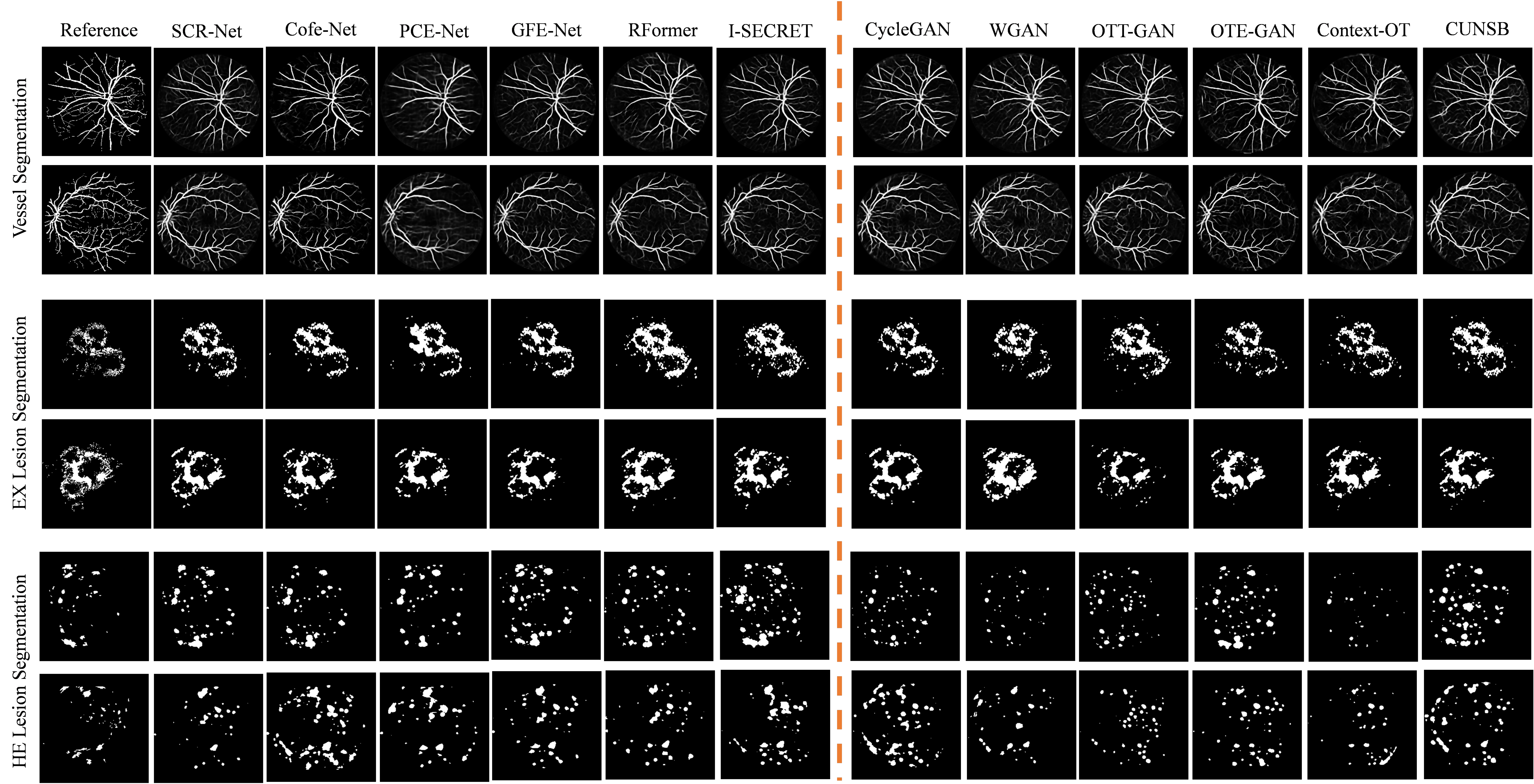}  
  \caption{ Illustration of Vessel and Lesion (EX and HE) Segmentation Experiments. The first column shows the reference segmentation masks, while the remaining columns display the segmentation results produced by all baseline models.
  }
  \label{fig:full-reference-segmentation}
\end{figure*}
\begin{figure*}[t]
  \centering
  \includegraphics[width=\textwidth]{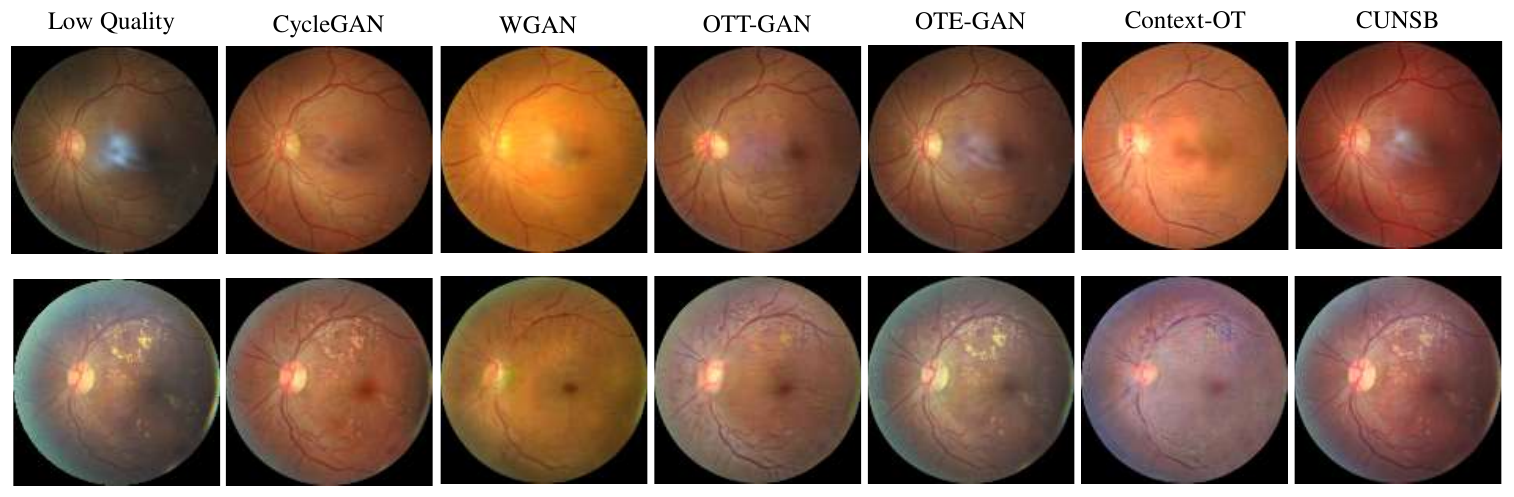}  
  \caption{ Illustration of the denoising results in the No-Reference Quality Assessment Experiments. The first column shows the input low-quality image, while the remaining columns display the synthetic high-quality images generated by all unpaired baseline models.
  }
  \label{fig:no-reference}
\end{figure*}

\section{No-Reference Quality Assessment Experiments Details}
We utilized the No-Reference Evaluation Dataset, including all unpaired baseline models for the No-Reference Assessment. These experiments evaluated the models' ability to learn and eliminate real-world noise. We maintained the experimental settings (e.g., hyperparameters) as outlined in Sec.~\ref{Sec:full-reference} to ensure a fair comparison.

\subsection{Lesion Segmentation}
Another U-Net model was employed for the downstream lesion segmentation task. The network consisted of 4 layers, with a base channel size of 64. The channel multiplier was set to 1 in the final layer and 2 in the remaining layers. The model was trained for 300 epochs using the Adam optimizer, with a batch size of 8. The initial learning rate was set to $2 \times 10^{-4}$, and a cosine annealing scheduler was applied, gradually reducing the learning rate to a minimum value of $1 \times 10^{-6}$. 

We utilized extensive data augmentation strategies to enhance model robustness. These included random horizontal and vertical flips, each with a probability of 0.5; random rotations with a probability of 0.8; random grid shuffling over $8\times 8$ grids with a probability of 0.5; and CoarseDropout, which masked up to 12 patches of size 
$20 \times 20$ to a value of 0, also with a probability of 0.5.

\subsection{DR grading.} We trained an NN-MobileNet model~\cite{deeplearning1} for the DR grading task using real-world high-quality images. The enhanced test images are used with the trained NN-MobileNet to infer DR grading classification. Enhancement performance is evaluated based on classification accuracy (ACC), kappa score, F1 score, and AUC. This evaluation primarily aims to assess whether the denoising model disrupts lesion distribution, potentially leading to inconsistencies with the original DR grading labels. 
During the training, we conducted 200 epochs with a batch size of 32 and an input size of $256 \times 256$. The AdamP optimizer was utilized with a $1 \times 10^{-3}$ weight decay and an initial learning rate of $1 \times 10^{-3}$. A dropout rate of $0.2$ was applied during training to mitigate over-fitting. Furthermore, the learning rate was dynamically adjusted using the Cosine Learning Rate Scheduler.

\subsection{Representation Feature Evaluation.}We employed two foundation models for fundus images, RetFound~\cite{zhou2023foundation} and Ret-CLIP~\cite{du2024ret}, to calculate the Fréchet Inception Distance (FID) between enhanced and real-world high-quality image feature representations. These metrics are referred to as \textit{FID-RetFound} and \textit{FID-CLIP}, respectively.

\textit{FID-RetFound} evaluates the preservation of disease-related information, while \textit{FID-CLIP} assesses the similarity of spatial structures and continuous features. To compute these metrics, the enhanced and real-world high-quality images were resized to $224 \times 224$ and normalized before being passed into the respective image encoders. The FID scores were then calculated based on the 1024-dimensional and 512-dimensional feature maps produced by RetFound and Ret-CLIP, respectively.

\subsection{Experts Annotation Evaluation.} To evaluate the quality of the enhanced images, we recruited six trained specialists to perform the manual evaluation. The evaluation criteria, as illustrated in Fig.~\ref{fig:expert-protocol}, included lesion preserving, background preserving, and structure-preserving. Each image was individually checked, and the results were carefully recorded. The six well-trained annotators conducted cross-evaluations on test images enhanced by different models to minimize subjective bias. Ophthalmologists further validated the final results to ensure accuracy and reliability.

\section{Result Illustrations}
We provide additional visualizations for all baseline models in the Full-Reference and No-Reference Quality Assessment Experiments. Specifically, Fig.~\ref{fig:full-reference-eyeq} presents the results of the Denoising Evaluation conducted on the EyeQ dataset. In contrast, Fig.~\ref{fig:full-reference-generalization} illustrates the Denoising Generalization Evaluation results on the DRIVE~\cite{drive} and IDRID~\cite{idrid} datasets. Fig.~\ref{fig:full-reference-segmentation} displays the outcomes of Vessel and Lesion (EX and HE) Segmentation. The results of the No-Reference quality assessment Experiments are outlined in Fig.~\ref{fig:no-reference}.
\end{document}